\documentclass[twocolumn, prd]{revtex4}
\usepackage[latin1]{inputenc}
\usepackage{amsthm,amsfonts}
\usepackage{amsmath}
\usepackage{amsthm}
\usepackage{amssymb}
\usepackage{appendix}
\usepackage{enumerate}
\usepackage{graphicx}

\begin{document}

\newtheorem{theorem}{Theorem}[section]
\newtheorem{Teo}{Theorem}
\newtheorem{lemma}[theorem]{Lemma}
\newtheorem{proposition}[theorem]{Proposition}
\newtheorem{corollary}[theorem]{Corollary}

\newenvironment{Proof}[1][Proof]{\begin{trivlist}
\item[\hskip \labelsep {\bfseries #1}]}{\end{trivlist}}
\newenvironment{definition}[1][Definition]{\begin{trivlist}
\item[\hskip \labelsep {\bfseries #1}]}{\end{trivlist}}
\newenvironment{example}[1][Example]{\begin{trivlist}
\item[\hskip \labelsep {\bfseries #1}]}{\end{trivlist}}
\newenvironment{remark}[1][Remark]{\begin{trivlist}
\item[\hskip \labelsep {\bfseries #1}]}{\end{trivlist}}

\newcommand{\Qed}{\nobreak \ifvmode \relax \else
      \ifdim\lastskip<1.5em \hskip-\lastskip
      \hskip1.5em plus0em minus0.5em \fi \nobreak
      \vrule height0.75em width0.5em depth0.25em\fi}

\title{Generalized Non-Commutative Inflation}
\author{U. D. Machado}
\email[]{machado@astro.iag.usp.br}


\author{R. Opher}
\email[]{opher@astro.iag.usp.br}

\affiliation{ Departamento de Astronomia, Geofísica e Ciências Atmosféricas, Universidade de São Paulo. \\ Rua do Matão, 1226-Cidade Universitária-São Paulo/SP- 05508-090}

\begin{abstract}
Non-commutative geometry indicates a deformation of the energy-momentum dispersion relation $f(E)\equiv\frac{E}{pc}(\neq 1)$ for massless particles. This distorted energy-momentum relation can affect the radiation dominated phase of the universe at sufficiently high temperature. This prompted the idea of non-commutative inflation by Alexander, Brandenberger and Magueijo (2003, 2005 and 2007). These authors studied a one-parameter family of non-relativistic dispersion relation that leads to inflation: the  $\alpha$ family of curves $f(E)=1+(\lambda E)^{\alpha}$. We show here how the conceptually different structure of symmetries of non-commutative spaces can lead, in a mathematically consistent way, to the fundamental equations of non-commutative inflation driven by radiation. We describe how this structure can be considered independently of (but including) the idea of non-commutative spaces as a starting point of the general inflationary deformation of $SL(2,\mathbb{C})$. We analyze the conditions on the dispersion relation that leads to inflation as a set of inequalities which plays the same role as the slow roll conditions on the potential of a scalar field. We study  conditions for a possible numerical approach to obtain a general one parameter family of dispersion relations that lead to successful inflation.

\end{abstract}

\maketitle

\section{Non-commutative spaces and deformed dispersion relations}

The fundamental concept in non-commutative spaces is the $C^{*}$-algebra 
$^{\underline{1}}$ \footnotetext[1]{$C^{*}$-algebra is a linear vector space $\mathcal{A}$ with an associative product $\cdot:\mathcal{A}\times\mathcal{A}\to\mathcal{A}$ (i.e. $(a\cdot b)\cdot c = a\cdot(b\cdot c) $); an operation called involution $*:\mathcal{A}\to\mathcal{A}$ that is defined  with the properties: $(A+B)^*=A^*+B^*$, $(\lambda A)^*=\overline{\lambda}A^*$, with $\lambda$ a complex number, $(AB)^*=B^*A^*$ and $(A^*)^*=A$ ;  a norm $||\mbox{ }||:\mathcal{A}\to\Re$  with respect to which the algebra is a Banach space (i.e. given a sequence $a_n$ of elements, if $\lim_{n\to\infty}||a_{n+m}-a_{n}||=0 $ for each $m>0$, then there exists an $a$ such that $\lim_{n\to\infty}||a_{n}-a||=0$ . );  the product is continuous with respect to the norm, i.e. $||AB||\leq||A||\cdot||B||$ , and the norm additionally  satisfies $||a^{*}a||=||a||^2$}.
The $C^{*}$-algebra is an algebraic idealization of a set of Hilbert space operators. Operations commonly defined on bounded Hilbert space operators, such as the product, linear combinations, adjoint and norm(defined as the  $||A||=\sup_{||u||\leq1}||Au||$) are defined on a $C^{*}$-algebra.   This set of operators can then be recovered from the formalism as a function only of the operations of the algebra.  On the other hand, when the product is commutative, we can alternatively represent it as the algebra of complex valued functions on some topological space. It is the content of the Gelfand-Naimark theorem that this correspondence is one to one.

 As opposed to the conventional differential geometric approach, in which we define the space and later the functions on it, we can define first the functions as an abstract algebraic entity, and implicitly specify the underlying space.

Non-commutative geometry generalizes the usual geometry by allowing the algebra to be non-commutative. By doing this, we lose the associated underlying space. Quantum mechanics is a case of a non-commutative space. It is completely defined by the specification of a $C^{*}$-algebra of observables, that replaces a commutative algebra of continuous functions of position and momentum (the commutative algebra of classical observables) with a non-commutative one. The rule to associate a non-commutative $C^{*}$-algebra to a commutative one is what defines a quantization procedure.

The idea that space-time coordinates should be replaced by non-commuting variables goes back to Heisenberg in an early attempt to regularize divergent integrals in quantum field theory \cite{Snyder}.  The success of the renormalization program ruled out this idea. Recent developments in string theory and M-theory suggest that non-commutative geometry could play a preeminent role in the physics at the Planck scale \cite{MTheory}, \cite{renormalization}.  Moreover, heuristically, one may expect that a final quantum gravity theory must incorporate some kind of uncertainty principle of space coordinates \cite{Noncommutative_Gravity}, since to localize arbitrarily a particle in space, according to quantum mechanics, requires arbitrarily high energy probing particles, which, according to general relativity, could create an event horizon over the measurement, invalidating it.

On the other hand, inflation has become a paradigm in cosmology, leading to various successful predictions. It is frequently argued in the literature, however, that we do not have as yet a realization of inflation based on fundamental physics and the usual realizations of inflation, based on weakly coupled scalar fields, have problems \cite{Inflation_Problems}.  This, by itself, is a motivation to consider alternative models of inflation based on new developments in physics beyond the standard model.

Quantum mechanics is a non-commutative version of classical phase space. Analogously, in the same way as quantum mechanics is not completely specified by commutation relations of phase space coordinates (although, by the Von Neumann theorem, all irreducible representations of the Heisenberg algebra are unitarily equivalent)  the non-commutative space is not uniquely defined by coordinate commutation relations. In quantum mechanics we use the correspondence principle to extend the quantization:
\begin{equation}[\hat{f},\hat{g}]=i\hbar\widehat{\{f,g\}}+\mathcal{O}(\hbar^2),\end{equation} where $f$ and $g$ are arbitrary classical functions of phase space, $\{f,g\}$ the Poisson bracket and $\hat{f}(q,p)$ indicates the quantum mechanical operator associated with the classical function $f(q,p)$ of phase space coordinates.

We use an additional rule to generate the $C^{*}$-algebra according  to the above quantization principle, the Weyl quantization, which is defined by the relations (given irreducible representations of the Heisenberg algebra):
\begin{eqnarray}f(q,p)=\frac{1}{(2\pi)^{\frac{n}{2}}}\int \tilde{f}(\mu,\nu)e^{i(\mu q+\nu p)}d\mu d\nu\\
\hat{f}(\hat{q},\hat{p})=\frac{1}{(2\pi)^{\frac{n}{2}}}\int \tilde{f}(\mu,\nu)e^{i(\mu \hat{q}+\nu \hat{p})} d\mu d\nu \\
e^{i(\mu \hat{q}+\nu \hat{p})}=e^{i\frac{\hbar\mu\nu}{2}}e^{i\mu\hat{q}}e^{i\nu\hat{p}}
\end{eqnarray}

Unfortunately, there exists infinitely many possible quantum algebras that we could associate with the commutative one  associated with flat space-time. We could postulate, for example, $[x^{\mu},x^{\nu}]=i\Theta^{\mu \nu}$, known as canonical non-commutativity, the most studied version of non-commutativity, where $\Theta^{\mu \nu}$ is defined as a constant antisymmetric quantity.  To find the associated $C^{*}$-algebra,  we use the Weyl quantization. We wish to define a quantum field theory on non-commutative space-time. The philosophy of a non-commutative geometry then says that only concepts formulated in terms of the algebra could be generalized for the non-commutative case.  Fortunately, quantum field theory is formulated in terms of a commutative algebra of classical fields. We then replace this algebra by the non-commutative one generated by the Weyl quantization under the general assumption $[x^{\mu},x^{\nu}]\neq0$.

This algebra can be represented as an algebra of Hilbert space operators, but if we could define an isomorphism between this operator algebra and the algebra of ordinary functions with a deformed product, i.e. $\hat{W}(f)\cdot\hat{W}(g)=\hat{W}(f\star g)$, we have a valid representation of the $C^{*}$-algebra that additionally maps the non-commutative theory into an ordinary field theory in commutative space-time. The $\star$ operation, known as the star product, is the basis for the most studied formulation of non-commutative field theory \cite{Douglas}. In other words, we map a non-commutative quantum field theory to an ordinary one with a deformed Lagrangian. We observe that the differential calculus defined by the same algebraic rules for smooth functions is not the same in the non-commutative space. A suitable definition of differentiation and integration must be studied case by case in order to define the action for non-commutative fields \cite{Douglas}:

\begin{eqnarray*}
 S=\int\left( \frac{1}{2}\partial_{\mu}\Phi\partial^{\mu}\Phi-\frac{m^2}{2}\Phi\Phi\right)d^4x\\
 \to S=\int \left(\frac{1}{2}\partial_{\mu}\Phi\star\partial^{\mu}\Phi-\frac{m^2}{2}\Phi\star\Phi\right)d^4x
\end{eqnarray*}

To see that this procedure may deform the usual relativistic dispersion relation of QFT at the free-field level(without the usual reference to quantum group formalism \cite{Lukierski}), we consider the k-Minkowski space-time, which is defined by the commutation relations $[x^i,x^j]=0$ and $[x^0,x^j]=\frac{i}{\kappa}x^{j}$ \cite{Camelia}, and the poles of the propagator (in Fourier space) for the free scalar field defined on it, which furnishes the dispersion relation:


\begin{equation}
\mathcal{C}_{\kappa}(k)-m^2=0
\end{equation}

where $C_{\kappa}$ is the deformed Casimir
\begin{equation}
\mathcal{C}_{\kappa}(p)=\left(2\kappa\sinh{\frac{k_0}{2\kappa}}\right)^2-\vec{k}^2 e^{-\frac{k_0}{\kappa}}
\end{equation}

The canonical non-commutativity does not lead to this kind of deformation at the level of free fields \cite{Douglas} and it is known that this kind of non-commutativity applied to scalar field driven inflation suffers from serious cosmological constraints \cite{PalmaCMB}. Consider additionally \cite{Gamboa} a different approach to non-commutative theories, which can be considered as a non-relativistic quantization scheme applied to a relativistic scalar field and which still leads to a deformed dispersion relation. Alternatively, consider the relationship between non-commutativity and two of the most important problems of cosmology today: The origin of large scale magnetic fields \cite{Bamba} and the cosmological constant problem \cite{RemoCC}.

The difference between considering $C^*$-algebras and the spectral triple approach for the non-commutative geometry of Connes \cite{Spectral} is that in spectral triple, not only the topology of space is codified in the algebra, but the metric field itself. Additional structures are included to accomplish this purpose. Spectral triple is therefore a generalization of the $C^*$-algebra to codify curved space. Consider the phenomenology associated to this approach \cite{ST1},\cite{ST2},\cite{ST3},\cite{ST4}, \cite{ST5}.

In section II we begin considering the role of the dispersion relation in the Wigner problem to represent symmetries. We then define the Hopf algebra concept and how it allows us to deform the Poincaré group in a more general way than that of a Lie group. We describe how the GNS construction of $C^*$-algebra theory can be used to map a Hopf algebra into something which has a quantum meaning. There are infinitely many non-equivalent representations of the Poincaré group. We propose a correspondence between the non-equivalent Hopf algebra representations and the Poincaré representations which corresponds to a particular known relativistic field theory. This correspondence fully defines the deformation of the quantum theory.  We propose a possible procedure to map a particular deformed dispersion relation into a full symmetry algebra which replaces Poincaré and realizes inflation (footnote 10). In section III we discuss the analogous slow-roll conditions of model and conditions on the dispersion relation which lead to successful inflation. In section IV, we propose a numerical algorithm for finding inflationary dispersion relations and comment on the qualitative differences on the generated spectrum in section V. In section VI we give our conclusions.                

\section{The approach of group theory for inflation}

The basic idea explored in \cite{Magueijo} is that the effect of generic models of non-commutative space-time can be codified in the associated modification of the energy-momentum relation affecting the calculation of the canonical partition function for radiation that in turn affects the early phases of the universe in thermal equilibrium.

This idea, we argue, is formalized in the Wigner approach to relativistic quantum theory, in which the basic problem is to construct representations of the Poincaré group as quantum symmetries without considering the quantization of a particular classical field. According to the Wigner theorem, a quantum symmetry can be extended from rays describing quantum states to the entire Hilbert space as a linear unitary or a anti-linear anti-unitary transformation.  For the proper and orthochronous part of the Poincaré group $P^{\uparrow}_{+}$ (for which the Lorentz subgroup satisfies $\Lambda_0^{ 0}>0$ and $\det\Lambda=1$), the problem reduces to construct unitary representations, since every group element is part of a one-parameter sub-group and thus it is the square of some other element. The square of a anti-linear anti-unitary operator is linear and unitary.   This problem summarizes into constructing the irreducible pieces by which every other representation can be constructed by direct sum (or integral). These basic parts are identified with the Hilbert space of one particle states. The dispersion relation $\mathcal{C}(p)$ is the fundamental information in this process because it is a (self-adjoint) function of space and time translation generators which commutes with all other generators of the symmetry group (Casimir of the Lie algebra) and defines a bounded operator$^{\underline{2}}$ \footnotetext[2]{An operator $A$ is bounded if $||A\Psi||\leq C ||\Psi|| $, with $C$ independent of $\Psi$} ($e^{iC(p)}$) which commutes with every element of the group.    By an infinite dimensional version of Schur`s lemma, a unitary representation of a group is irreducible if and only if every bounded operator which commutes with every element of the group is a multiple of the identity.

There is an infinite number of possible representations that can be constructed by knowing the irreducible representations. But we want to consider a pre-inflationary thermodynamic universe for which it suffices to know the representation of a free field.

To define the Hilbert space representation of the symmetry group of a free field, we define first the $N$-particle representation as the symmetrized (or antisymmetrized) $N$-fold tensor product $(U^{\otimes N}_{\lambda})_{S,A}$ 
(where $\lambda$ is a set of indexes which labels irreducible representations, the Casimir eigenvalue among them, and $S$, $A$ stands for the symmetrized or antisymmetrized product respectively), which acts on $N$-fold tensor product of the one particle Hilbert spaces $\left(\mathcal{H}^{\otimes N}_{\lambda}\right)_{S,A}$.
We then define the Hilbert space representation as the direct sum of all $N$-particle representations  
\begin{equation}\label{Fock}
U=\sum_{N=0}^{\infty} {}^{\oplus}\left(U^{\otimes N}_{\lambda}\right)_{S,A}
\end{equation}
 which is defined on 
 \begin{equation}
 \mathcal{H}=\sum_{N=0}^{\infty}{ }^{\oplus}\left(\mathcal{H}^{\otimes N}_{\lambda}\right)_{S,A}
 \end{equation}
 ($N=0$ corresponds to vacuum trivial representation). For each $\mathcal{H}_{\lambda}$ we choose a base of common eigen-states of Momentum, Hamiltonian and the Casimir operator $\Psi_{p,\sigma}$ (every base element associated with the same eigenvalue of Casimir and allowing additional degrees of freedom in $\sigma$. The number of them is associated with the dimensionality of irreducible representations of group elements which leaves the four-momentum invariant $^{\underline{3}}$ \footnotetext[3] {This is the Little group. For the Poincaré's group and massive particles this group is $SO(3)$, which is a compact group. For a compact group, there are at most a countable number of irreducible representations inequivalents, all finite dimensional. For zero mass, the group is $ISO(2)$, which is not compact, leading to the existence of irreducible representations of infinite dimension. Additional conditions are needed to select the finite dimensional representations}). It is nothing more than the Fock space constructed as a symmetry representation problem rather than a quantization of a classical field.

To calculate the canonical partition function for radiation confined in a cubic box, we need a Hilbert space representation of the Hamiltonian ($Z(\beta)=Tr\left(e^{-\beta {H}}\right)$), but it is diagonal in this base and additionally the Momentum is quantized by imposing periodic boundary conditions on the unitary representations of space translations due the walls of the cubic box ($U(x)=U(x+L)$).

The above analysis follows directly (except for questions about intrinsically projective representations) for a deformation at high energies of the Lie group $P^{\uparrow}_{+}$  and its associated Lie algebra, which we denote as $\mathfrak{p}$, more precisely, its universal enveloping algebra $\mathcal{U}(\mathfrak{p})$. The universal enveloping algebra is an associative algebra, i.e., elements of $\mathfrak{p}$ can now be multiplied by using an associative product, in such a way that the Lie bracket is represented as a commutation relation. This allows $ [X ^ {i}, X ^ {j}] = F (X ^ {k}) $, where $ F (X ^ {k}) $ is an analytic function of the generators of the Lie algebra $X^{k}$. $F(X^{i})$ is well defined (i.e., unambiguous), provided it can be decomposed as linear combination of ordered products of generators.

The reason for this deformation is that the Casimir is a function of commutation relations between generators. If the Casimir changes, then the commutation relation between generators necessarily changes.  Moreover, if we deform the structure constants of the Poincaré Lie algebra as functions of energy-momentum which converge to the original values for the low energy-momentum limit, then, the deformed algebra of generators, acting on  one-particle states (in the energy-momentum representation) whose support is restricted to low values of energy and momentum, is indistinguishable from the action of the Poincaré Lie algebra  (here, $C^{ij}_k = C^{ij}_k(H,P)$ as a function of operators, but by hypothesis $[H, P] = 0$, then in the energy-momentum representation $C^{ij}_k(H,P)$ is an ordinary function of real values):

\newcommand{\act}{\rhd}
\begin{gather} \label{lowEnergyLim}
\Psi=\int d\mu(p)\phi(p)\Psi_p\\ \nonumber
  \mbox{supp}\{ \phi(p)\}\in\{E<E_{max},P^{j}<p^{j}_{max}\}\\ \nonumber
 \Rightarrow [X^i,X^j]\act\Psi= iX^kC^{ij}_k(H,P)\act\Psi\to i X^k{C^{ij}_k}^{(0)}\act\Psi\mbox{,}
\end{gather}
 where $\mbox{supp}$  denotes the support, i.e, the closed set outside of which the four-momentum function $\phi $ is zero. $C^{ij}_k(H, P)$ converges to the undeformed Poincaré structure constant (${C^{ij}_k}^{(0)}$) when $E_{max}$ and $p^i_{max}$ approaches zero. Here, $\act$  denotes the action of  generators on one-particle states and $ \to $ denotes the convergence in the strong sense (i.e. $||X^k C^{ij}_k(H,P)\act\Psi-  X^k {C^{ij}_k}^{0}\act\Psi||\to0$) $^{\underline{4}}$ \footnotetext[4]{This limit is always valid in a realization of the algebra as bounded operators. When the generators are at least self-adjoint, or $||X^kC^{ij}_k(H,P)\act\Psi-  X^k{C^{ij}_k}^{0}\act\Psi||\to0$ or is not a convergent sequence.}. This illustrates the physical argument that the typical eigenvalues of the generators can effectively change the commutation relations.

\subsection{The Hopf algebra description}

Since $[X^{i},X^{j}]\neq C^{ij}_{k}X^{k}$, for constant $C^{ij}_{k}$, the group is no longer a Lie group (i.e. there is no associated group manifold), but we still have a classical group. However, this deformed algebra can be considered in the formalism of universal enveloping Hopf algebras, or quantum enveloping algebras, related to the idea of quantum groups. The Hopf algebra is a useful concept, since it covers many concepts of group theory, such as finite groups, Lie groups and Lie algebras, into a single common structure. It represents a set of transformations which act not only on vector spaces but on general algebras like the $ C^{*} $-algebra, with the fundamental difference that not all transformations are invertible (they have an inverse in a weaker sense called antipode).

To define the Hopf algebra $H$, we need to specify its action on the product of elements of the algebra: $X\act (f\cdot g)=\sum_{ij} (X_{(i)}\act f)\cdot(X_{(j)}\act g)$, where the rule $\Delta : {H}\to{H}\otimes{H}$ given by  $\Delta: X\to\sum_{ij}X_{(i)}\otimes X_{(j)}$ is called a coproduct. When one of the multiplying elements is the unit, it must satisfy $X\act (1\cdot g)=\sum_{ij} (X_{(i)}\act 1)\cdot(X_{(j)}\act g)=\sum_{ij}\epsilon(X_{(i)})\cdot({X_{(j)}\act g})=X\act g$, where the rule $\epsilon: X\to\mathbb{C}$  is called a counit. A Hopf algebra has a generalized notion of inverse called antipode with the properties $\cdot (S\otimes id)\Delta h=\epsilon (h)$. The motivation for this notion is how a group $G$ acts on itself by the adjoint representation: $g\act a=gag^{-1}$, in such way that $g\act(a\cdot b)=(g\act a)\cdot (g\act b)$, which implies that $\Delta g=g\otimes g$. The antipode is then defined with the properties $h\act (1\cdot b)=h_{(i)}1S(h_{(j)}) \cdot h_{(i)}bS(h_{(j)})=\epsilon(h)\cdot h\act(b)$, where the adjoint action of $H$ on itself is defined as $h\act b=h_{(i)}bS(h_{(j)})$ (we used the shorthand notation $ h_{(i)}\otimes h_{(j)}$ for $\sum_{ij} h_{(i)}\otimes h_{(j)}$), see for example \cite{Majid} for a complete reference. The algebra in which the Hopf algebra acts according to the above rules is called a $H$-module algebra.

For a Lie algebra, for example, $\Delta X=X\otimes 1+ 1\otimes X$, $\epsilon(X)=0$ and $S(X)=-X$, which states that the elements of the Lie algebra act as derivations (i.e., obeying the Leibnitz rule) on an algebra of differentiable functions with the usual commutative product $(f\cdot g)(x)= f(x)\cdot g(x)$.  This algebra can represent the states of a particle in the energy-momentum representation and illustrates how we can add an algebraic structure on the Hilbert space $L^{2}(\mathbb{R}^{N})$ without affecting the predictions of quantum theory encoded in the linear structure. Furthermore, Hopf algebras are suitable for describing the symmetries of non-commutative spaces, since they define the action of symmetries on non-commutative $C^{*}$-algebras as in \cite{HopfAlgebra}.

To connect Hopf algebras, which describe the non-commutative symmetries (but not only), with quantum theory in Hilbert space, we must study their realizations. The key point is that exactly as for Lie algebras the knowledge of the structure constants is sufficient to construct representations (the adjoint representation, for example), the main advantage of describing the deformation of the enveloping algebras as Hopf algebras is that its structure contains enough information to realize it. To do so, we must consider the dual $H^{*}$ of the Hopf algebra $H$, which are the linear functionals on it: $\left<\phi,h\right> \to \mathbb{C}$, $\phi\in H^{*}$, $h\in H$. This is a Hopf algebra with structure induced by the one in $H$: $\left< \phi\psi,h\right>=\left< \phi\otimes\psi,\Delta h\right>$, $\left< \Delta\psi,g\otimes h\right>=\left< \psi,g\cdot h\right>$, $\left< 1,h\right>=\epsilon(h)$ and $\epsilon(\psi)=\left<\psi,1\right>$.

We define the left coregular action $R^*$ of a Hopf algebra $H$ on its dual $H^*$ turning it into a  $H$-module algebra:  $\left<R^*_{g}(\phi), h\right>=<\phi,hg>$. We then define an involution operation on $H^*$, analogous to the involution operation on $C^{*}$-algebra, which will work as the adjoint operation (it is not unique) and satisfy $X=X^*$ $^{\underline{5}}$ \footnotetext[5] { We specify the antilinear involution operation $*: H\to H$, i.e. the adjoint operation, compatible with the Hopf algebra structure: $(\Delta h)^{*}=(\Delta h)^{*\otimes*}$, $\epsilon(h^*)=\overline{\epsilon(h)}$, $(S\circ*)^2=id$ which specifies the Hopf $*$-algebra}. Finally, we use the GNS construction of $C^{*}$-algebras which realizes it as operators on Hilbert space.

The main point of the GNS construction is that each state $\mathcal{S}$ (i.e., positive linear functional, which means that $S(A^*A)\geq0$) induces a representation of the $C^{*}$ algebra as operators in Hilbert space such that $S(A)=\left<0\right|A\left|0\right>$, where $\left|0\right>$ is a cyclic vector, i.e., operators of the algebra  acting on $\left|0\right>$ generate all the physical states (more precisely, it generates a subspace which is dense in Hilbert space).

The Hopf $H^*$ algebra has its natural ``vacuum state" given by its left integral, which is defined by $\int\phi=TrL_{\phi}\circ S^{2}$  (here, $Tr$ $^{\underline{6}}$ \footnotetext[6]{The trace of an abstract $C^*$-algebra is uniquely defined, except by normalization, by linearity and cyclic property} is a cyclic trace, $L_{\phi}$ is the left action of the algebra on itself $L_{\phi}\psi=\phi\psi$, $\circ$ is the composition of linear operators and $S$ is the antipode). The integral of the Hopf algebra is the analogue of the integration of functions defined on the group manifold of the Lie group with a measure which is invariant by the action of the group. Thus, the generators of the Hopf algebra can be represented as self-adjoint generators of one-parameter subgroups of unitary transformations (under the condition $X^{*}=X$).

In the GNS construction, the states are $\Psi=\psi$, $\psi\in H^*$ and if $\Psi_1-\Psi_2=\phi$ such that $\int \psi^*\cdot\phi=0$ for all $\psi\in H^*$, then $\Psi_1=\Psi_2$ $^{\underline{7}}$ \footnotetext[7]{There are technical details on the faithfulness and irreducibility of the representation. The norm in the algebra $C^*$ implies that the operators are bounded and the domain is the entire Hilbert space. When we remove the norm, the operators have a common dense domain which can not be extended to the whole Hilbert space.}. The left coregular action of $H$ on $H^*$ respects the above equivalence relation, and therefore defines a linear operator in Hilbert space: $ \int \psi^*\cdot\left( h\act(\psi_1+\phi)\right)=\int (h^*\act\psi)^*\cdot\psi_1=\int \psi^*\cdot (h\act\psi_1)$ $^{\underline{8}}$ \footnotetext[8]{ The $\int$ operation can always be adjusted such that $\int \psi^*\cdot h\act\phi=\int (h^*\act \psi)^*\cdot\phi$ }. The inner product is $(\Psi,\Phi)=\int\psi^*\cdot\phi$. The product on $H^*$ also respects the equivalence relation and therefore induces a product in Hilbert space. If $H$ is cocommutative, i.e. $\Delta(H)=H_{(1)}\otimes H_{(2)}=H_{(2)}\otimes H_{(1)}$ then the product in $H^*$ is commutative. This is the case of Lie algebras but is not the case in non-commutative spaces. Usually the name quantum group is reserved for Hopf algebras which are not commutative or cocommutative.

According to Wigner, irreducible representations of symmetry groups are related to particles. There are infinitely many irreducible non-equivalent representations.  The representation above is however very special, it is the faithful one in Hilbert space, every other can be derived from it.  Indeed, states (every  bounded  linear functional) corresponds to representations and reciprocally, provided that the associated representation has a cyclic state. A representation is irreducible if and only if the associated normalized state ($\mathcal{S}(1)=1$) cannot be decomposed in the convex linear combination form: $\mathcal{S}=\lambda\mathcal{S}_1+(1-\lambda)\mathcal{S}_2$ for $\lambda\in(0,1)$ and normalized states $\mathcal{S}_1$ and $\mathcal{S}_2$. Every representation with a cyclic state $\Psi$ is unitarily equivalent to its associated GNS construction. In irreducible representations every state is cyclic. Given a faithful representation in Hilbert space, every other state is given by $\mathcal{S}(A)=Tr(\rho A)$ where $\rho$ is a density matrix. More exactly, every state is given by the weak* closure of these states $^{\underline{9}}$
 \footnotetext[9]{Given a sequence of states $\mathcal{S}_n$, if $\mathcal{S}_n(A)$ is convergent for every element $A$ of $C^*$-algebra, the limit $\lim \mathcal{S}_n$ is an state}. Therefore, every irreducible representation can be obtained by a second GNS construction applied to a faithful representation in Hilbert space, since we know the general form of the states.

The approach outlined by us is sufficiently general because it includes the case of quantum enveloping algebras, which describes the symmetries of noncommutative spaces \cite{HopfAlgebra}, and Lie algebras (which is a special case of Hopf enveloping algebras). And it is  more general, since we have no proof (at least one that is known by the authors) that every deformation of the dispersion relation of interest to cosmology is associated with some type of non-commutative space. However, it is not difficult to prove that any possible deformed dispersion relation that can be written as $E/f(E)=p$  can be realized (in a way not unique) as the Casimir of some deformed enveloping algebra $^{\underline{10}}$
\footnotetext[10] {In fact, consider $X_i$ the infinitesimal generators of homogeneous Lorentz group. They are contravariant vectors, or first order differential operators. Consider the diffeomorphism $\Phi$ given by $\bar{E}=E$ ; $\bar{p}=p/f(E)$. Define the new algebra as multiplication operators $E$, $p$ and $\bar{X}_i=\Phi_{*}X_i$, the pushfoward operator $\Phi_{*}X_i\act f=X_i \act f\circ \Phi$. This new algebra satisfy $[E,p]=0$; $[\bar{X}_i,\bar{X}_j]=[X_i,X_j]$, but $[\bar{X}_i,E]=(\bar{X}_i\act E)=F(E,p)$ and $[\bar{X}_i,p]=(\bar{X}_i\act p)=G(E,p)$; As required $[f(E,p),E]=[f(E,p),p]=[f(E,p),\bar{X}_i]=0$, where $f(E,p)=f^2(E)p^2-E^2$. Moreover, the involution condition ${X_i}^{*}=X_i$ is compatible with the algebra.}.

\subsection{Basic equations of the model}
Non-commutative radiation is then a perfect fluid characterized by a pressure and energy density associated to the trace of the representation (\ref{Fock}), and defined by the equations \cite{Magueijo}:

\begin{equation}\label{rho}
   \rho(E,T)=\frac{1}{\pi^2}\frac{E^3}{\exp{E/T}-1}\frac{1}{f^3}\left|1-\frac{Ef^{'}}{f}\right|
   \end{equation}
   \begin{equation}\label{p}
   p=\frac{1}{3}\int \frac{\rho(E,T)}{1-\frac{Ef^{'}}{f}}dE
   \end{equation}
   \begin{equation}\label{rhoT}
    \rho=\int\rho(E,T)dE
   \end{equation}
   \begin{equation}\label{dispersion}
   E^2=p^2f^2
   \end{equation}
where we take $ c=k_B=\hbar=1$. The inflation model discussed in \cite{Brandenberger} was defined by the choice:

   \begin{equation} \label{f(E)}
   f=1+(\lambda E)^{\alpha}
   \end{equation}

This is equivalent to defining a more general type of non-relativistic quantum theory on a local inertial reference frame of FRW space-time, after which we calculate the energy momentum tensor at thermodynamic equilibrium and extend this object to curved space time by conventional equivalence principle.

The calculation has implicitly the hypothesis that the number of photon internal degrees of freedom does not change. For more general deformations of relativistic symmetries it does not change the equation of state $w=p/\rho$, since it changes the pressure and energy density by a multiplicative factor. There is the possibility that the early universe has a symmetry different from that at low energy and in a discontinuous phase transition retrieves the Poincaré invariance of conventional physics. This is, for example, the case of a variable speed of light cosmology \cite{MagueijoVSLC}, which can be considered a breaking of local Lorentz symmetry, which does not lead to inflation, but it can solve the horizon problem, among others, for example.

\subsection{Interacting Generalization}

As a final remark, to obtain the equations previously considered in \cite{Magueijo}, we have worked with the simplest possible realization of a deformed enveloping algebra in Hilbert space, which can be regarded as a deformation of the free QFT defined on the inertial reference frame (\ref{Fock}).  How can we generalize the above prescription to a more complex interacting field theory like $\phi^4$, for example?

The usual physical paradigm tells us to somehow modify the associated action. We could consider an alternative prescription which has the advantage of avoiding by construction any consideration about renormalization: the perturbative quantization scheme applied to usual relativistic QFT give us, order by order, the N-point correlation functions $\left<0\right|\phi(x_1)\cdots\phi(x_N)\left|0\right>$. The  Wightman reconstruction theorem and related schemes, which are  based on the GNS construction described earlier, allows us to recover from N-point correlation functions, not only a representation for fields in Hilbert space (as operator valued distributions), but an associated Poincaré group representation in Hilbert space. Since the first Casimir of the Poincaré Lie algebra has a continuous spectrum, a fundamental result in the theory of group representation in  infinite dimensional Hilbert space allows us to express it as the direct integral:

\begin{equation}
\mathcal{U}^{\mathcal{P}}=\mathcal{U}_0\oplus\int^{\oplus} d\mu(\lambda)\sum_{\sigma}^{\oplus}\nu(\lambda,\sigma)\mathcal{U}_{\lambda,\sigma}^{\mathcal{P}}
\end{equation}
which acts on Hilbert space:
\begin{equation}
\mathcal{H}=(c\Psi_{0})\oplus\int^{\oplus} d\mu(\lambda)\sum_{\sigma}^{\oplus}\nu(\lambda,\sigma)\mathcal{H}_{\lambda,\sigma}^{\mathcal{P}}
\end{equation}
where $\mathcal{U}_{\lambda,\sigma}^{\mathcal{P}}$ defines an irreducible representation of the Poincaré group which are labeled by Casimir eigenvalue $\lambda=m^2$ and a discrete index $\sigma$  (the spin or helicity). $\mathcal{U}_0$ is the trivial vacuum representation and $\nu(\sigma,\lambda)$ is a degenerence of the representation (i.e. a number of times that it appears in the decomposition) which can be finite or infinite. $c\Psi_{0}$ is the one-dimensional space generated by the vacuum $\Psi_{0}$. It means that at each $\lambda$ value and at each set of $\sigma$ values, we define a direct sum of $\nu(\lambda,\sigma)$ copies of a Hilbert space $\mathcal{H}_{\lambda,\sigma}^{\mathcal{P}}$ in which an irreducible representation acts. The inner product of this Hilbert space is an integration with respect to some (unique) measure $d\mu(\lambda)$ of the inner product of all spaces.

We then postulate, as a possible definition, the following deformation:
\begin{equation}
 \mathcal{U}_{\lambda,\sigma}^{\mathcal{P}}\to\mathcal{U}_{\lambda',\sigma'}^{\mathcal{NP}}
\end{equation}
here $\mathcal{U}_{\lambda,\sigma}^{\mathcal{NP}}$ is the irreducible representation of the non-relativistic deformed Hopf enveloping algebra which recovers $\mathcal{U}_{\lambda,\sigma}^{\mathcal{P}}$ in the low energy-momentum limit as in (\ref{lowEnergyLim}). In other words, the rule to obtain the correspondence $\lambda'(\lambda,\sigma)$ and $\sigma'(\lambda,\sigma)$ is such that there exists a unitary mapping $\pi$ from states with Poincaré irreducible realization to irreducible realization of the deformed algebra whose matrix elements agree in the low energy-momentum limit:
\begin{equation}
\left(\pi(\Psi_{\lambda,\sigma}),\mathcal{U}^{\mathcal{NP}}_{\lambda',\sigma'}\pi(\Phi_{\lambda,\sigma})\right)\to\left(\Psi_{\lambda,\sigma},\mathcal{U}_{\lambda,\sigma}^{\mathcal{P}}\Phi_{\lambda,\sigma}\right)
\end{equation}
 It is possible to prove, by the Lebesgue dominated convergence theorem $^{\underline{11}}$ \footnotetext[11]{Lebesgue's theorem says that if $g_N(x)\to g(x)$ for all $x$ and $|g_N(x)|\leq f(x)$ for some integrable $f(x)$, then $g(x)$ is Lebesgue integrable and $\int g_N(x)\to\int g(x)$. Since we know that for original QFT there exists Hilbert space vectors $\Psi_{\sigma,\lambda}$ in each $\mathcal{H}_{\lambda,\sigma}$ such that the inner product integral converges, and the unitary transformation satisfies $\left|\left(\pi(\Psi_{\lambda,\sigma}),\mathcal{U}^{\mathcal{NP}}_{\lambda',\sigma'}\pi(\Phi_{\lambda,\sigma})\right)\right|\leq||\Psi_{\lambda,\sigma}||\cdot||\Phi_{\lambda,\sigma}||\leq||\Phi_{\lambda,\sigma}||^2+||\Psi_{\lambda,\sigma}||^2$ which is a mensurable function with finite integral, we have the fulfillment of conditions of the Lebesgue's theorem.  }, that the associated deformed unitary algebra weakly converges(i.e., in the sense of matrix elements: $\left(\pi(\Psi),\mathcal{U}^{\mathcal{NP}}\pi(\Phi)\right)\to\left(\Psi,\mathcal{U}^{\mathcal{P}}\Phi\right)$ )  to the associated low energy Poincaré representation if all intermediate states of direct integral decomposition converge as (\ref{lowEnergyLim}), which corresponds to states of sufficiently low energy and momentum.

\section{Conditions on $f(E)$ analogous to the Slow-Roll conditions on the scalar field potential}

\subsection{Acceptable thermodynamic and low energy limit conditions}

 Many phenomenological models of Trans-Planckian physics are based on deformations of the energy-momentum relations (see for example: \cite{Doubly}, \cite{Amelino_Gamma}, \cite{PalmaStringy}, \cite{RemoDR}, \cite{BrandenbergerTransP} and \cite{BrandenbergerTransP2}). As we do not have a consensus on the correct non-commutative version of space-time, or the high energy deformation of Poincaré symmetry, we could consider the cosmological consequences associated with the correspondent deformed energy-momentum relation and thereby, in principle, put cosmological constraints on physical principles beyond the standard model.

We know from the scalar field realizations of inflation, that it is not the specific form of the potential that leads to inflation, but the validity of the slow roll conditions,  $\left|\frac{V_{,\phi\phi}}{V}\right|<<1$ and $\left(\frac{V_{,\phi}}{V}\right)^2<<1, $ in the limit of large field values, for models like hybrid inflation and chaotic inflation. These conditions imply that there exists a great variety of initial conditions, in field configuration space, which can produce the right amount of inflation \cite{Review_Inflation}.

We could ask if the same situation occurs in non-commutative inflation: What are the conditions on $f(E)$ that lead to successful inflation in the homogeneous limit (by homogeneous, we mean without considering constraints in the generated perturbation spectrum which is considered in \cite{Brandenberger_Perturbations} and \cite{Brandenberger_Evolution_Perturbations} for $f(E)=1+(\lambda E)^{\alpha}$). By successful, we mean it produces a minimum e-folding number and does not have a graceful exit problem.

The choice $f=1+ (\lambda E) ^\alpha$ made in \cite{Brandenberger} is not as arbitrary as may seem at first sight. Indeed, the substitution of this relation in the denominator of  Eq. (\ref{p}) leads to a constant equation of state $w =p/\rho$ in the limit of high $T$ if the spectrum of $\rho (E,T)$ attains its maximum at ever increasing values of energy for ever increasing values of temperature in such a way that the greatest contribution of the integral of Eq. (\ref{rho}) comes from regions with arbitrary high energies  as in the case of the usual Planck spectrum. This approximation is the justification for the choice.

As shown numerically in \cite{Brandenberger}, the hypothesis of peaks of $\rho(E,T)$ for higher and higher values of temperature fails, however, for the choice of $f(E)$ made in the inflationary range.  For high temperatures we have instead a saturated peak (See Fig 1 in \cite{Brandenberger}).

\begin{figure}[h]
  \includegraphics[width=8cm]{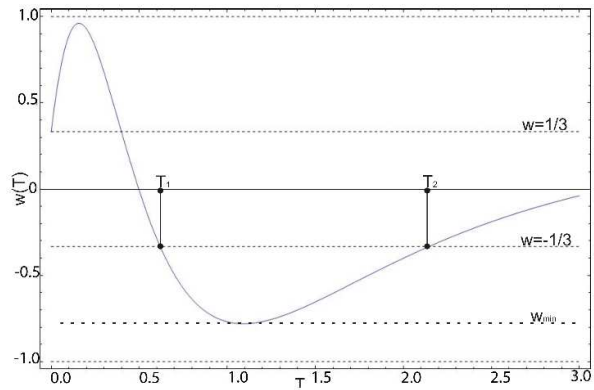}
  \caption{Generic graphic of $w(T)$ versus $T$. }\label{Generic_Omega_T}
\end{figure}

We may expect that for a general $f(E)$ we have a generic curve for $w$ as shown in Fig.\ref{Generic_Omega_T}, with $w\sim 1/3 $ for low temperatures and an inflation period $-1\leq w\leq-1/3$ for some high temperature interval.

In equation (\ref{rho}), $\frac{E^2}{f^3}\left|1-\frac{Ef^{'}}{f}\right|=\left|p^2\frac{dp}{dE}\right|=\left|\frac{1}{3}\frac{d(p^3)}{dE}\right|$ is proportional to the density of one particle states from $\Omega(E)dE=\Omega(p)dp$. From this, we observe the first physical requirement: some choices of a dispersion relation lead to divergent expressions of energy density and pressure at thermal equilibrium, because it leads to too much particle states per energy interval. We require additionally that $f(E\to0)\to1$ i.e. for low energy, the dispersion relation is usual.  Additionally we require that $w(T\to0)=1/3$ for the conventional low energy equation of state. These conditions are related, and  with additionally $\lim_{E\to0}Ef'(E)=0$ we have the first set of conditions for the dispersion relation(see Appendix A for the proof).

\begin{Teo}
If $\frac{1}{f^3}\left|1-\frac{f'E}{f}\right|\leq C(1+E^k)$  and $\frac{1}{f^3}\leq C'(1+E^{k'})$ for some real positive constants $C$ and $C'$ and some integers $k$ and $k'$, $f(E)$ and $f'(E)$ continuous for $E\geq 0$ with $\lim_{E\to0}f(E)=1$, and $\lim_{E\to0}Ef'(E)=0$ then the expressions for the energy density and pressure at thermal equilibrium are finite and:
 $$\lim_{T\to 0}\frac{p(T)}{\rho(T)}=\frac{1}{3}$$
\end{Teo}

\subsection{Minimum e-fold number condition}

The other condition that needs to be fulfilled by inflation is that we have a  minimum e-folding number $\mathcal{N}$. The estimate of this number depends on the problem in consideration (flatness, horizon, etc) \cite{Review_Inflation}. The number is $\gtrsim 60$.

 From Fig. \ref{Generic_Omega_T} we can estimate that: 
 \begin{equation}
 \frac{1}{2}\ln{\frac{\rho (T_2)}{\rho (T_1)}}<\mathcal{N}<\frac{1}{3(1-w _{min})}\ln{\frac{\rho (T_2)}{\rho (T_1)}}.
 \end{equation}
 This estimate comes from the conservation of energy equation that leads to $-3(1-\frac{1}{3})d\ln{a}<d\ln{\rho}<-3(1-w _{min})d\ln{a}$. We thus come to the conclusion that for a minimum e-folding number $\mathcal{N}$,  it suffices that $\frac{\rho (T2)}{\rho (T1)}>\exp{2\mathcal{N}}$. In particular, we can assume $\frac{\rho (T2)}{\rho (T1)}\to \infty$. This is the case for inflationary models with a constant equation of state in the high energy limit.

 Let us now obtain the conditions to assure a constant equation of state in the high temperature limit (see Appendix B for the proof. Additionally, the conditions of Theorem 1 are automatically satisfied by the following conditions):
\begin{Teo}\label{limit_w}
Define $g=1-\frac{f^{'}E}{f}$ with the following properties:
\begin{enumerate}
\item $g(E \to 0)=1$ and g is continuously differentiable for $E\geq 0$.
\item There exists a finite number $N$ of energies $E_{0 1}, E_{0 2},...E_{0 N}$ such that $g(E_{0 j})=0$ and $\left|\frac{dg(E_{0 j})}{dE}\right|>0$.
\item There exists an $\epsilon >0$ such that $g(E)\leq -\epsilon $ for $E\geq E_1$
\end{enumerate}
Under such conditions, we have

\begin{multline} \label{WinftyS_1}
\lim_{T \to \infty} w (T)=\frac{1}{3}\Bigl(\int_0^{E_{0 1}}E^2/f^3dE-\int_{E_{0 1}}^{E_{0 2}}E^2/f^3dE+...\\
+\int_{E_{0 N-1}}^{E_{0 N}}E^2/f^3dE-\int_{E_N}^{\infty}E^2/f^3dE\Bigr) \Bigl/ \int_0^{\infty}\left| g \right|E^2/f^3dE
\end{multline}

where each  integral involved converges.
\end{Teo}

It is important to comment here that in \cite{Brandenberger} the approximation
\begin{equation}
\lim_{T \to \infty}w (T) \approx \frac{1}{3(1-\alpha)}
\end{equation}
was made instead.

Let us adopt the following convention:
\begin{definition}
The set $G_N$ consists of all g functions satisfying all conditions of Theorem \ref{limit_w} and the functions have $N$ roots.
\end{definition}

The first condition is related with the low energy limit for the dispersion relation. The zeros of the $g(E)$ function are local maximums and minimums of momentum as a function of energy ($\frac{dp}{dE}=0$ ) i.e., points of transition between positive and negative pressure modes (if $\left|\frac{dg}{dE}\right|>0$). Inflation necessarily needs negative pressure, $\frac{dp}{dE}<0$ is then needed because the mechanism of inflation is such that when the universe expands the wave-length of one-particle states increases, momentum diminishes in inverse proportion and for particles with a conventional dispersion relation in which $\frac{dp}{dE}>0$ it implies that energy diminishes, but for particles with $\frac{dp}{dE}<0$ energy actually increases leading to negative pressure (according \cite{Brandenberger}). 

  The existence of $\frac{dp}{dE}=0$ points is then needed for inflation and implies that we must have a non-invertible function $p(E)$.  The number of such points is related with more complicated oscillations of the equation of state as a function of temperature. The condition (3) is the important condition here, because it implies $p(E\to\infty)=0$, which in turn leads to the existence of a maximum momentum below which every energy level is mapped, which we can associate with a minimum probable scale.

The number density of momentum eigenstates for radiation in a box is a function only of the periodic boundary condition of the continuous unitary representation of space translations: if  $U(x+L)=U(x)$ and $\left<\Psi\right|U(x)\left|\Phi\right>$ is continuous  for every $\left|\Psi\right>$ and $\left|\Phi\right>$, then 
\begin{equation}U(x)=\sum_N e^{\frac{2\pi i N x}{L}}\frac{1}{2\pi}\int_{0}^{L} e^{\frac{-2\pi i N x}{L}}U(x)dx=\sum_N e^{\frac{2\pi i N x}{L}} E_N,
\end{equation}
 where $E_N$ are mutually orthogonal projections, such that $U(x)=e^{-iPx/\hbar}=e^{-i\sum_N \frac{2\pi\hbar N }{L} E_N x/\hbar}$.

A limitation for momentum, in principle, leads to a limitation of the number of one particle states which can be occupied by photons, but, for this model, we interpret non invertible functions $p(E)$ as allowing more than one energy branch for each momentum, leading to arbitrary more particle states, even if momentum is limited. This interpretation is actually rigorous since we represent the Hilbert space of one-particle states as (N-component) functions $\Psi_{\sigma}$ defined on the deformed mass shell $\mathcal{C}(p)-m^2=0$ whose modulo is square (Lebesgue) integrable with respect to some symmetry invariant measure concentrated on the mass shell and consistent with positive energy condition. This measure can be written, for example, as $\delta(\mathcal{C}(p)-m^2)\theta(p^0)d\mu(p)$ where $d\mu(p)$ is a measure over momentum space, other than $d^4p$, which is invariant under the deformed action of the group in momentum space.  Condition 2 leading to a finite number of energy branches and condition 3 leading to $p(E\to\infty)=0$ and in turn to maximum momentum, which implies that the total number of particle states is finite. We actually need only $p(E\to\infty)<\infty$, but the zero value has another function related with the entropy.

Because, additionally, the function $\frac{E}{T}n(E,T)=\frac{E/T}{e^{E/T}-1}$ has the right convergence properties (uniformly bounded and uniformly convergent to unity in each compact interval, according to a more rigorous proof in the appendix)  we have then a convergent expression for $\rho/T$ in the high temperature limit. Every one particle state contributes to total pressure with a term ${-\partial E}/{\partial V} $ which depends on $\frac{dp}{dE}$, condition 3 additionally  implies that momentum tends to zero so fast that we analogously have a convergent expression for $\frac{p}{T}$ in the high temperature limit leading to a constant equation of state in the high energy limit. As a consequence, this conditions leads to a constant entropy density $s=\frac{\rho+p}{T}$ in the high energy limit. The condition of a finite number of one particle states alone is not sufficient to lead to a maximum attainable entropy because we do not have the particle number conservation constraint limiting the number of possible microstates configurations.

The $\left|\frac{dg}{dE}\right|>0$ condition is used here only to assure that the $g$ function crosses the zero line and not only touches it. Below,  it is critical for the continuity theorem. Physically, it means that we have a local maximum and minimum for momentum, not saddle points.

As a consequence of the above analysis, the conditions imply that the thermal spectrum saturates and attains a temperature independent shape proportional to $|g(E)|E^2/f^3$ (always having zeros leading to a multi-peak partner) multiplied by temperature.   We have then the high energy Stephan-Boltzmann law: $\frac{d\ln \rho}{d\ln T}\to1$ when $T\to\infty$. This situation is numerically obtained in \cite{Brandenberger} for an interval of the parameter $\alpha$ that can be shown to satisfy all the stated requirements (See Fig 2 in \cite{Brandenberger}).

\subsection{Graceful exit condition}
The problem of graceful exit comes from the fact that if in some neighborhood of $T$ we have  $ \rho (T)$ a differentiable invertible function, we can write $w = w (\rho)$, the equation of state for an isentropic fluid.  For a FRW metric, $T_{;\nu}^{\nu \mu}=0$ becomes $\frac{d(\rho)}{\rho}=-3(1+w)\frac{da}{a}$ and we have that if there exists a $T_0$ in this neighborhood such that $\rho _0=\rho (T_0)$ and $w ( \rho _0)=-1$ , the equation above could be written as $\frac{d\ln{\rho}}{d\ln{a}}=-3(1+w (\ln{\rho}))$, and has a unique solution  $\rho (\ln{a})=\rho _0$  for   $\rho (\ln{a_0})=\rho _0$

 From the inverse function theorem, we have that if $ \rho (T)$ is a  continuously differentiable function of $T$ and for some value $T_0$  we have $\left| \frac{d\rho}{dT}\right|>0$, then we have $ \rho (T)$ an invertible function in  a neighborhood $\mathcal{U}$ that contains $T_0$ .This means that if there exists such a temperature $T_0$ then at this temperature the Universe never exits the de-Sitter phase. As we show below (the proof is in appendix C), $\frac{d\rho}{dT}>0$ always.
 
We might believe from the above that all dispersion relations leading to $\lim_{T \to \infty}w\in (-1/3,-1)$ lead to an acceptable cosmology. This is not true. We must have $w>-1$ to avoid the graceful exit problem as discussed above. It is not true in general that $\lim_{T\to \infty}w>L$ implies $w(T)>L$ for all $T$.

Consider, for example, the function $\theta(E,E_0,\delta,\alpha)$ $^{\underline{12}}$ \footnotetext[12]{ It has a value $0$, when $E<E_0$; value $\alpha$, when $E>E_0+\delta$; and a  value $\alpha\int_{-1}^{-1+2\frac{E-E_0}{\delta}} e^{-\frac{1}{1-x^2}}dx/I$, with $I=\int_{-1}^{1} e^{-\frac{1}{1-x^2}}dx$, when $0<E-E_0<\delta$. It is a smooth transition between two constant values.}. It is an infinitely differentiable function that we use to build the following $g$ function:
\begin{equation}\label{_THETA_}
\begin{split}
g(E)&=1+\theta(E,0,1,-1.05)+\theta(E,100,3,3)\\
    &\quad+\theta(E,200,3,-3.2)
\end{split}
\end{equation}

\begin{figure}[h]
  \includegraphics[width=8cm]{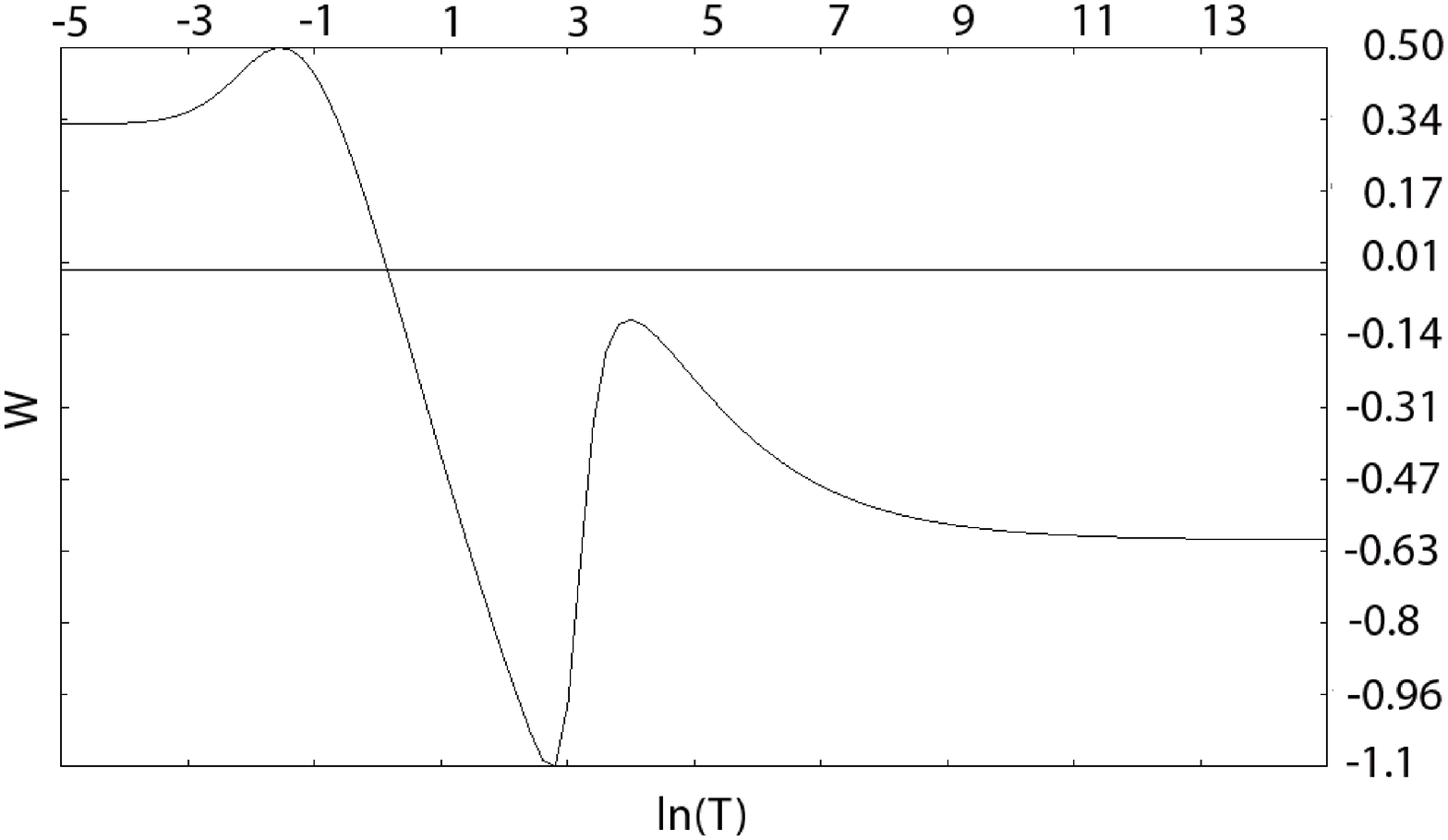}
  \caption{w(T) versus ln(T) for the $g$ function of Eq. \ref{_THETA_}. We have in this example a high temperature equation of state in the inflationary range ($w(T\to\infty)\approx-0.6$), but for some T, we have $w(T)<-1$ leading to a graceful exit problem. }\label{Gold_Graphic}
\end{figure}

This function corresponds to a momentum that first decreases with energy toward zero, then it increases and then decreases again towards zero. It produces the graph in Fig. \ref{Gold_Graphic} that shows a graceful exit problem: We have at temperature $T$  that $w(T)=-1$ leading to a graceful exit problem, even with the  asymptotic equation of state in the inflationary range.

We are led to the next question:  What conditions assure  $w(T)>-1$ for all $T$. We assure this by the following theorem (see Appendix C for the proof):

\begin{Teo}
if $g\in G_1$, once we have $\lim_{T \to \infty }w>-1$, we will necessarily have $w(T)>-1$ for all $T$ if $g>-1/3$.
\end{Teo}
 This is a sufficient condition to avoid  the graceful exit problem and not a necessary one. We suspect that the number of zeros of $g(E)$ is the relevant feature and that the theorem is valid without the restriction $g(E)>-1/3$.  
\section{The numerical algorithm}

It is not obvious how to invert Eq. (\ref{WinftyS_1}) in the interval $(-\frac{1}{3},-1)$ to identify generic deformations of the dispersion relation that lead to an inflationary high temperature equation of state. Our strategy will be to postulate a one parameter ($\alpha$) family of dispersion relations such that we are assured by construction that there will be an interval of the parameter $\alpha$ where inflation takes place successfully.

We do this by building a sequence of functions $g_N$, $ N=1,2,...$ compatible with all the requirements considered so far in such a way that $\lim_{N\to \infty}w_N^{\infty}<-1$ ($\lim_{T\to\infty}w_N(T)\equiv w_{N}^{\infty}$) and another sequence such that $\lim_{N\to \infty} w_N^{\infty}>-1/3$. We continuously interpolate with a $\alpha$ parameter between two sufficiently advanced elements of these sequences (in such a way that the chosen elements satisfy the same inequality limits), in a way compatible with the previous requirements and such that the $\lim_{T\to \infty} w$ depends continuously on the $\alpha$ parameter.

One of these sequences can be constructed using the following theorem and we will have $lim_{N\to \infty}w_N^{\infty}<-1$ for sufficiently large $N$ (see Appendix D for the proof):

\begin{Teo}
Suppose a functional sequence $g_N$, $N=1,2\cdots$ such that ($g_N\in G_1$, $g_N>-\frac{1}{3}$ and $g_N(\xi_N)=0$):
\begin{enumerate}[(I)]
\item $g_N(E)\geq 1-{E}/{\xi_N}$, $0\leq E\leq\xi_N$ for all $N$.\\
\item $|g_N(E)|<\epsilon_N$ for $E\geq\xi_N$.\\
\item $g_N<c_N$, $0<c_N<c$ and $(c_N-1)\xi_N<k$ for $E<\xi_N$.\\
\item $\epsilon_N\to 0$ when $N\to\infty$.\\
\item $\exists$ $\xi'$,  $0<\xi'<\xi_N$ such that $g_N<1+\alpha E$, $\alpha>0$ and $E\leq\xi'$\\
\end{enumerate}
Then, $\lim_{N\to\infty}w_{N}^{\infty}=-\infty$.
\end{Teo}

For building the other sequence, we use the following proposition and we will assure $lim_{N\to \infty}w_N^{\infty}>-1/3$ for sufficiently large $N$ (see appendix E for proof):

\begin{Teo}
Suppose a functional sequence with $g_N(E)$ $($$g_N(E)\in G_1$ with $g_N(E)>-1/3$ and $g_N(\xi_N)=0$$)$ satisfying:
\begin{enumerate}[(I)]	
\item $g_N(E)<1+\alpha E$ for $E<\xi'$  and for all $N$ and $0<\alpha<\infty$\\
\item $g_N(E)>1-\beta E$ for some  $0<\beta<\infty$, for $E<\xi''<\xi'$ and for all $N$\\
\item $g_N(E)<1/3-\epsilon$, for some $\epsilon>0$, $E\geq\xi'$ and for all $N$.\\
\item $\xi_N\to\infty$ when $N\to \infty$ \\
\item $g_N(E)<-1/3+\epsilon_N$ for $E>\xi_N+\Delta_N$, with $\epsilon_N\to 0$ and $\Delta_N\to0$ when $N\to \infty$\\
\end{enumerate}
, then, for sufficiently large $N$ we have $w_N^{\infty}>0$
\end{Teo}

The above results still work without the restriction $g_N>-\frac{1}{3}$ , but our results are not sufficient to assure the graceful exit from inflation.

By interpolation we mean a function $g(\alpha,E)$ with $\alpha \in [\alpha_{1},\alpha_{2}]$ such that $g(\alpha_{1},E)=g_{N}(E)$ with $w_N^{\infty}>-1/3$  and $g(\alpha_{2},E)=g^{*}_{N}(E)$ with ${w^{*}}_N^{\infty}<-1$, $g(\alpha,E)$ continuous (i.e. $||(\alpha_1,E_1)-(\alpha_2,E_2)||<\delta$ implies $|g(\alpha_1,E_1)-g(\alpha_2,E_2)|<\epsilon$ for any $\epsilon$ and some $\delta$) $g(\alpha,E)\in {G_1}$ for all $\alpha$.

Extremely important for our procedure to be valid, in general, is that we can assure continuity of the high temperature equation of state with respect to the parameter of interpolation $\alpha$.
We must be careful, in general, while building a one parameter family of continuous functions and expecting that the integral of these functions is continuous with the associated parameter. Consider the example involving an integral on a unbounded interval (exactly as in Eq. (\ref{WinftyS_1}) )

$$g(x,\alpha)=\left\{\begin{array}{rc}
\frac{\alpha}{\sqrt{\pi}}e^{-{(\alpha x)^2}}\quad\mbox{if}\quad\alpha>0\\
0\quad\mbox{if}\quad\alpha=0
\end{array}\right.
$$

We have that $g(x,\alpha\to 0)=0$, but $\lim_{\alpha\to0}\int_{-\infty}^{\infty}g(x,\alpha)dx=1$ and $\int_{-\infty}^{\infty}g(x,0)dx=0$. Then, we do not have continuity of the integral with respect to the $\alpha$ parameter. The definition of the class $G$ of functions makes it easy to assure the desired continuity, but even in Fig. 3 of \cite{Brandenberger} we have an example of the discontinuity of a high temperature equation of state for this model (see Appendix F for the proof):

\begin{Teo}
Suppose $g(E,\alpha)$ a limited $(|g(E,\alpha)|<C$ for all $\alpha$ and all $E)$ and differentiable function such that $g(E,\alpha)\in G_1$ for all $\alpha\in[\alpha_1,\alpha_2]$ $($and as a consequence of definition of the $G_1$, $\frac{dg(E,\alpha)}{dE}<0$ when $E=\xi_N(\alpha)$ such that $g(E,\xi_N(\alpha))=0$ $)$, then:
\begin{equation}\label{goal}
\frac{1}{3}\frac{\int_0^{\xi_N(\alpha)}E^2/f(E,\alpha)^3dE-\int_{\xi_N(\alpha)}^{\infty}E^2/f(E,\alpha)^3dE}{\int_0^{\infty}\left| g(E,\alpha) \right|E^2/f(E,\alpha)^3dE}
\end{equation}
is continuous with respect to the $\alpha$ parameter.
\end{Teo}

The above last three theorems only state that we can give an almost arbitrary initial guess for $g(E)$ and, by controlling a finite number of deformation parameters of this initial guess, in a progressive sequence, we can construct a one parameter family of deformations of the energy-momentum relation that will contain an inflationary range of asymptotic equations of state (i.e. $w(T\to\infty)$ will contain in its range the $(-1/3,-1)$ interval), it will have a conventional low energy equation of state and will not have a graceful exit problem. The Functional Sequence Theorems give flexibility to deform the curve along all of its extension. Provided that we have found the extremes of our family, say $g_1(E)$ with $w(T\to\infty)>-1/3$ and $g_2(E)$ with $w(T\to\infty)<-1$ if $g_1(\xi_1)=0$ and $g_2(\xi_2)=0$ with $\xi_1<\xi_2$ and, $dg_1(E)/dE<0$ and $dg_2(E)/dE<0$ in the $(\xi_1,\xi_2)$ interval, then $\alpha g_1(E)+(1-\alpha) g_2(E)$ with $0\leq\alpha\leq1$ satisfy all the requirements of the continuity theorem, for example.

\section{considerations about generated perturbations}

\begin{figure}[h]
  \includegraphics[width=8cm]{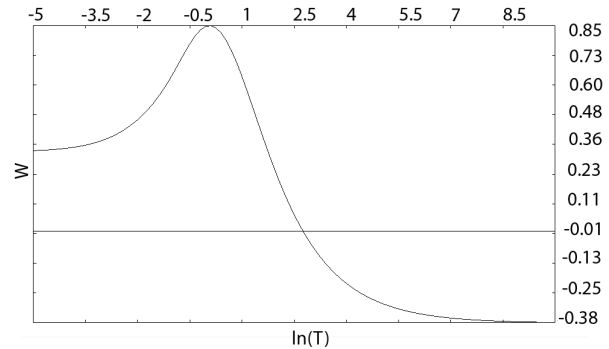}\\
  \caption{$w(T)$ versus $\ln(T)$ for $f=1+(\lambda E)^{\alpha}$}\label{Branden}
\end{figure}

\begin{figure}[h]
  \includegraphics[width=8cm]{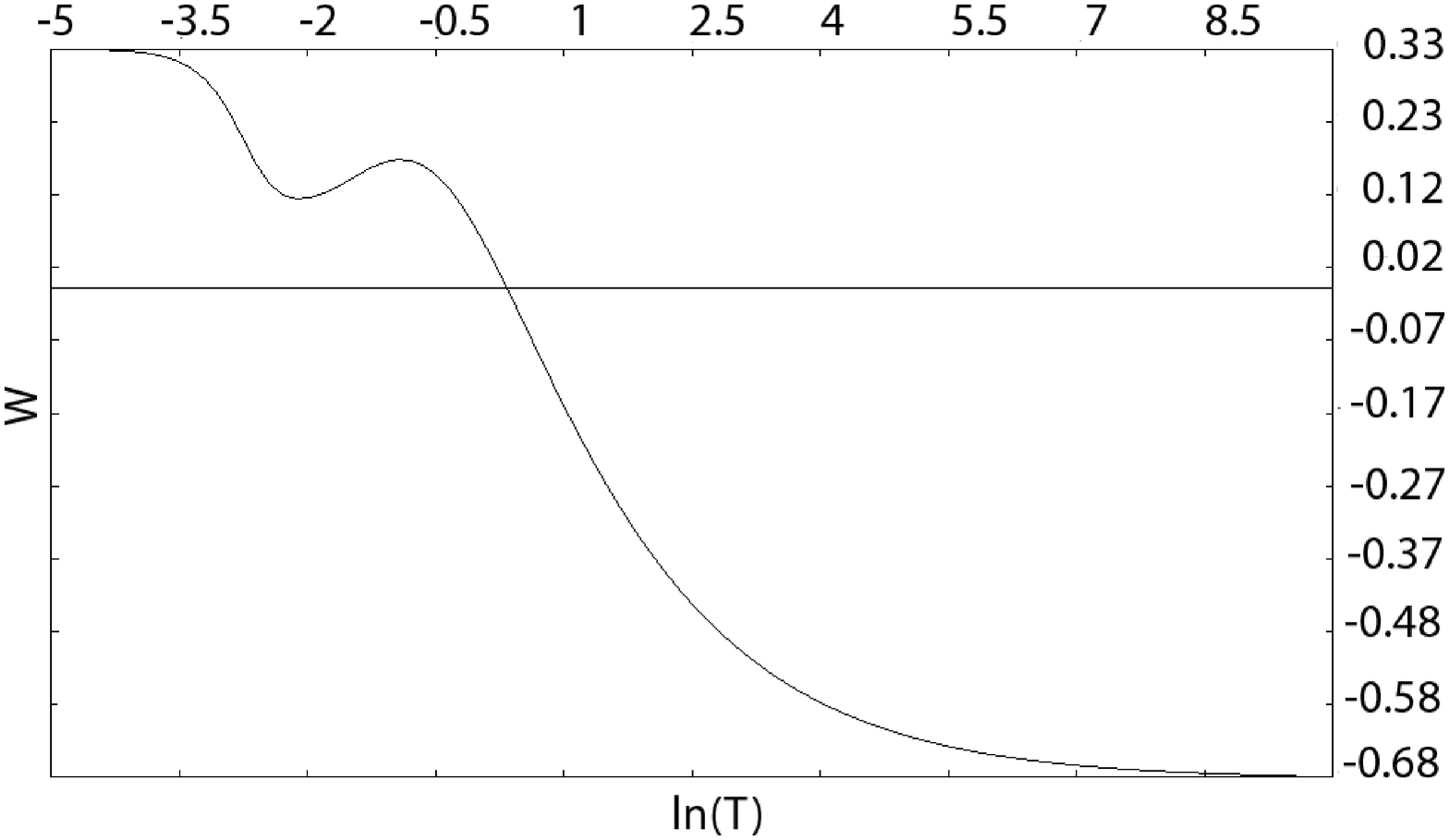}\\
  \caption{$w(T)$ versus $\ln(T)$ for $g(E)=1+\theta(E,0,1,2)+\theta(E,1,2,-3.2)$}\label{OUR}
\end{figure}

 We may compare the equations of state $w(T)$ for the family $f=1+(\lambda E)^{\alpha}$ studied in \cite{Brandenberger} with alternative choices given by, for example, $g(E)=1+\theta(E,0,1,2)+\theta(E,1,2,-3.2)$ with $\theta$ defined previously by Eq. (\ref{_THETA_}) shown in Fig. \ref{Gold_Graphic}. These two curves are shown in Figs (\ref{Branden}) and (\ref{OUR}). Fig (\ref{Branden}) shows the general behavior of the equation of state $w$ of $f=1+(\lambda E)^{\alpha}$: There is a monotonic increase until a maximum greater than $1/3$ and then it decreases monotonically to the inflationary range. Alternative choices as in Fig (\ref{OUR}), describe more general behaviors in which we can have arbitrary oscillations of the equations of state. The graphs show the fact that the difference between the choices satisfying the conditions considered is the transition between the two asymptotic limits of equation of state.

Since we have a constant asymptotic equation of state followed by a radiation dominated phase, the analysis done in \cite{Brandenberger_Perturbations} is applicable. There, the quantum number fluctuation of the variable $v$, defined by $\Phi=4\pi G\sqrt{\rho+\mathcal{P}}\frac{z}{k^2c_s}\left(\frac{v}{z}\right)'$, is given by the Bose-Einstein distribution at the moment that scales cross the thermal correlation length $T^{-1}$. The scales then freeze when crossing the sound horizon.

For the range of scales that leave the horizon at sufficiently high temperatures so that the equation of state is arbitrarily close to its asymptotic value and reenter the horizon in the radiation dominated phase in which $w=1/3$ (the limit of large wavelengths), the power spectrum is unaffected by different choices of the dispersion relation that satisfy the conditions considered by us. In the small wavelength limit, however, different choices of $f(E)$ lead to modifications. In fact, as stated earlier, the situation is analogous to inflation with scalar field satisfying the conditions of slow rolling, which generally provides the power spectrum almost scale invariant in the limit of large wavelengths, but leads to a red tilt in the limit of small wavelength, whose magnitude depends on the specific choice of potential.

It is known that different choices of slow roll potentials may lead to an overproduction of primordial black holes or dark matter small halos, for example, exactly as for different choices of $f(E)$. Moreover, the evolution of scales which reenter the horizon at earlier times (or leave it later) is modified for different choices of the dispersion relation of energy-momentum, leaving its own imprint. In this regime, the density perturbations experience variations of the speed of sound that affect the horizon scale and, consequently, its linear growth phase. For perturbation generation in this model see   \cite{Brandenberger_Perturbations}, with the posterior linear evolution between the two asymptotic values of the equation of state
for $f(E)=1+(\lambda E)^\alpha$  considered in \cite{Brandenberger_Evolution_Perturbations}. The details of the CMB imprint of specific $f(E)$ solutions deserves a separate study.

\section{Conclusion}

In this paper, we consider a more rigorous formulation of non-commutative inflation. Instead of following the usual physical point of view which is to deform the action to take into account all possible additional effects, we explore the Wigner point of view in which the basic problem of relativistic field theory is represent the Poincaré group according to the Wigner theorem. In this framework, the dispersion relation is the basic ingredient of the representation problem. Here we consider that non-commutative space is a mathematical object called a $C^*$-algebra. And, as pointed in literature \cite{HopfAlgebra}, there exist a mathematical object called a Hopf algebra which can be interpreted as a set of transformations which act on generic algebraic structures respecting their structures. Therefore, it was suggested that this object can codify the symmetry structure of non-commutative spaces. This object was, however, previously used only as an intermediate step to deform the field action. 

 We use, instead, a basic theorem of the $C^*$-algebra theory called GNS construction to direct connect a Hopf algebra with quantum theory. We propose a prescription to deform the Poincaré group representation associated to a free and interacting relativistic field theory which transforms it into a inflationary representation, that is, one which can lead to inflation at high energies, without leading to infinities, or violations of unitarity and assuring a suitable low energy limit. We do it based on the Wightman reconstruction theorem and a direct integral form of unitary group representation in infinite dimensional Hilbert space.  This prescription, applied for free fields, leads to the usual equations of non-commutative inflation. In the case of interacting fields, it shows us the inflationary character of general representations of suitable algebras. The algebra that puts everything to work must have certain properties. We propose an algebra that replaces Poincaré and realizes non-commutative inflation (footnote 10). This algebra, although not related to a given non-commutative space, is not a linear Lie algebra related to a Lie group, but a Hopf algebra.

Moreover, we study, based on previously obtained equations of non-commutative inflation, the ``slow roll conditions for the dispersion relations of the model" that assure a minimum e-folding number, graceful exit and acceptable thermodynamic behavior, including a correct low energy limit for the equation of state. Previous authors only considered one particular choice of the dispersion relation, but a good cosmology theory avoids fine tuning. Our findings put the non-commutative inflation in analogy with scalar field inflation in which it is not the specific form of the potential that leads to inflation, but the validity of the slow-roll conditions.  

We find that the high temperature behavior of an acceptable model must be the same as the particular example studied by previous authors, but the exit from inflation must be different until the low energy limit is attained. It can affect the linear evolution of perturbations and must be numerically analyzed. It implies that the long wavelength limit of the power spectrum must agree for all viable models, exactly in the same way as the slow-row conditions generically predicts a scale invariant power spectrum for long wavelength, and leads to a red tilt at small wavelength.    The details of perturbation theory must be calculated numerically. Additional constraints come from big-bang-nucleosynthesis, which marks the energy scale before which all non-standard effects should disappear. Therefore it is the energy scale for which the dispersion relation must be the usual one.   Considering a numerical analysis, we propose a numerical algorithm for finding solutions according to previous slow roll conditions. The proofs for these conditions are somewhat cumbersome and are left as appendices.

\begin{acknowledgements}
U.D.M. thanks the Brazilian agency CNPQ (142393/2006-1) for financial support. R.O. thanks FAPESP (06/56213-9) and the Brazilian agency CNPQ (300414/82-0) for partial support. We would like to thank professor Marcelo Gomes and his group for interesting and helpful discussions.
\end{acknowledgements}

\appendix
\section{Proof of theorem 1}
\numberwithin{equation}{section}

\begin{theorem}

If $\frac{1}{f^3}\left|1-\frac{f'E}{f}\right|\leq C(1+E^k)$  and $\frac{1}{f^3}\leq C'(1+E^{k'})$ for some real positive constants $C$ and $C'$ and some integers $k$ and $k'$, $f(E)$ and $f'(E)$ continuous for $E\geq 0$ with $\lim_{E\to0}f(E)=1$, and $\lim_{E\to0}Ef'(E)=0$ then the expressions for energy density and pressure at thermal equilibrium are finite and:
 $$\lim_{T\to 0}\frac{p(T)}{\rho(T)}=\frac{1}{3}$$

\end{theorem}
\begin{Proof}
$$\rho(T)=\int\frac{1}{\pi^2}\frac{E^3}{e^{E/T}-1}\frac{1}{f^3}\left|1-\frac{f'E}{f}\right|dE $$
and,
$$p(T)= \frac{1}{3}\int\frac{1}{\pi^2}\frac{E^3}{e^{E/T}-1}\frac{1}{f^3}\mbox{sign}(g(E))dE\mbox{,}$$
Where $$\mbox{sign}(g)=\left\{\begin{array}{rc}
         1\mbox{ if $g(E)\geq 0$ }\\
        -1\mbox{ elsewhere}\end{array}\right.\mbox{,}$$
where $g(E)=1-\frac{f'E}{f}$

Since $f(E)$ is continuous, $\lim_{E\to0}f=1$ and $\lim_{E\to0}|Ef'(E)|=0$ ($f'$ is continuous on $E\geq0$) there exists $\delta>0$ such that $1-\epsilon\leq\frac{1}{f^3}\left|1-\frac{f'E}{f}\right|\leq1+\epsilon$ if $E\leq\delta$ for each $\epsilon>0$. Then, rewrite $\rho(T)$ as:
$$\rho(T)=\int_0^{\delta}\rho(E,T)dE+\int_{\delta}^{\infty}\rho(E,T)dE$$
We show that: $$\lim_{T\to0}\frac{\int_0^{\delta}\rho(E,T)dE}{\int_{\delta}^{\infty}\rho(E,T)dE}=\infty$$
 for each $\delta$ and a similar result for $p(T)$:
\begin{equation*}
\begin{split}
&\quad\lim_{T\to0}\frac{\int_0^{\delta}\rho(E,T)dE}{\int_{\delta}^{\infty}\rho(E,T)dE}\geq\lim_{T\to0}\frac{\int_0^{\delta}\frac{E^3}{e^{E/T}-1}(1-\epsilon)}{\int_{\delta}^{\infty}\frac{E^3}{e^{E/T}-1}C(1+E^k)}\\
&\geq\lim_{T\to0}\frac{\int_0^{\delta}\frac{E^3}{e^{E/T}-1}(1-\epsilon)}{\int_{\delta}^{\infty}\frac{E^3}{e^{E/T}}C(1+E^k)}=\lim_{T\to0}\frac{T^4\int_0^{\delta/T}\frac{y^3}{e^{y}-1}(1-\epsilon)dy}{\int_{\delta}^{\infty}\frac{E^3}{e^{E/T}}C(1+E^k)}\\
&=\lim_{T\to0}\frac{T^4\int_0^{\delta/T}\frac{y^3}{e^{y}-1}(1-\epsilon)dy}{e^{-\delta/T}P(\delta,\frac{1}{T})}=\infty
\end{split}
\end{equation*}
where $P(\delta,\frac{1}{T})$ denotes a polynomial of finite degree on variables $\delta$ and $\frac{1}{T}$

Then:
\begin{equation*}
\begin{split}
&\lim_{T\to0}\frac{1}{3}\frac{ \int_0^{\delta}\frac{E^3}{e^{E/T}-1}(1-\epsilon)}{ \int_0^{\delta}\frac{E^3}{e^{E/T}-1}(1+\epsilon)}\leq\lim_{T\to0}\frac{p(T)}{\rho(T)}\\
&\leq \lim_{T\to0}\frac{1}{3}\frac{ \int_0^{\delta}\frac{E^3}{e^{E/T}-1}(1+\epsilon)}{ \int_0^{\delta}\frac{E^3}{e^{E/T}-1}(1-\epsilon)}
\end{split}
\end{equation*}

since $\epsilon$ is arbitrary, we conclude the proof.\Qed
\end{Proof}
\section{Proof of theorem 2}
\begin{theorem}

Define $g=1-\frac{f^{'}E}{f}$ with the following properties:
\begin{enumerate}
\item $g(E \to 0)=1$ and g is continuously differentiable for $E\geq 0$.
\item There exists a finite number $N$ of energies $E_{0 1}, E_{0 2},...E_{0 N}$ such that $g(E_{0 j})=0$ and $\left|\frac{dg(E_{0 j})}{dE}\right|>0$.
\item There exists an $\epsilon >0$ such that $g(E)\leq -\epsilon $ for $E\geq E_1$
\end{enumerate}
Under such conditions, we have

\begin{multline} \label{WinftyS}
\lim_{T \to \infty} w (T)=\frac{1}{3}\Bigl(\int_0^{E_{0 1}}E^2/f^3dE-\int_{E_{0 1}}^{E_{0 2}}E^2/f^3dE+...\\
+\int_{E_{0 N-1}}^{E_{0 N}}E^2/f^3dE-\int_{E_N}^{\infty}E^2/f^3dE\Bigr) \Bigl/ \int_0^{\infty}\left| g \right|E^2/f^3dE
\end{multline}

where each  integral involved converges.

\end{theorem}
\begin{Proof}

Let us show a particular case in which $N=1$ and $E_{0 1}=E_{0}$. There is no additional work in showing the general case.  The condition $\left|\frac{dg(E_{0 j})}{dE} \right|>0$  implies, by the inverse function theorem, that in a neighborhood of $E_{0 j}$, $g(E)$ is a diffeomorphism, in particular injective, such that for each value of $E$ in this neighborhood such that $|g(E)|>0$ there is no other $E$ with the same $g(E)$ value. It implies that the $g(E)$ curve crosses the zero line in this neighborhood.

The first condition assures that we can write $f$ as $\exp{\int_0^E\frac {1-g(x)}{x}dx}$, because $\frac {1-g(x)}{x}$ has a definite limit in $x \to 0$ given by $\left| g^{'} (0)  \right|$.
It produces an $f$ continuously differentiable and $f(0)=1$.

The first step is to show that each integral involved is finite: The first integral $\int_0^{E_0}E^2/f^3dE$ is finite because $E_0$ is finite and by construction $f > 0$. The second integral $\int_{E_0}^{\infty}E^2/f^3dE$ is finite because, by (3), the $E_1$ must be greater than $E_0$ and we have that  $\frac{f(E \geq E_1)}{f(E_1)} > \exp{\int_{E_1}^{E}\frac{1+\epsilon}{x}dx}=\exp{[(1+\epsilon)(\ln{E}-\ln{E_1})]} =CE^{1+\epsilon}$
that implies $\frac{E^2}{f^3}<\frac{E^2}{f^3(E_1)C^3E^{3+3\epsilon}}$ for $E \geq E_1$, that, in turn, leads to $\int_{E_1}^{\infty}\frac{E^2}{f^3}<\frac{1}{C^3f^3(E_1)}\int_{E_1}^{\infty}\frac{1}{E^{1+3\epsilon}}dE=
\frac{1}{C^3f^3(E_1)}\frac{E_1^{-3\epsilon}}{3\epsilon}<\infty$. The integral $\int_{E_0}^{E_1}E^2/f^3dE$ is finite for the same reason as the first integral.
The last integral $\int_0^{\infty}\left| g \right|E^2/f^3dE$ is finite because $\int_0^{\infty}\left| g \right|E^2/f^3dE=\int_{0}^{\infty}\frac{1}{3}\left|\frac{dp^3}{dE}\right|dE$ and  $\int_{E_1}^{\infty}\frac{1}{3}\left|\frac{dp^3}{dE}\right|dE=-\int_{E_1}^{\infty}\frac{1}{3}dp^3/dE=\frac{1}{3}p(E_N)^3$, because condition 2 implies that $p=E/f(E)\to0$ when $E\to\infty$.

We then can write
\begin{equation*}
\begin{split}
 w(T)&=\frac{1}{3}\frac{\int_0^{\infty} \frac{\rho (E,T)}{1-f^{'}E/f}dE}{\int_0^{\infty} \rho(E,T)dE} \\
     &=\frac{1}{3}\frac{\frac{1}{T}\int_0^{\infty} \frac{\rho (E,T)}{1-f^{'}E/f}dE}{\frac{1}{T}\int_0^{\infty} \rho(E,T)dE}\\
     &=\frac{1}{3}\frac{\int_0^{E_0}\frac{\zeta(E,T)}{T}\frac{E^2}{f^3}dE-\int_{E_0}^{\infty}\frac{\zeta(E,T)}{T}\frac{E^2}{f^3}dE}{\int_{0}^{\infty}\frac{\zeta(E,T)}{T}\frac{E^2}{f^3}\left|g\right|dE}\\
\end{split}
\end{equation*}
where $\zeta(E,T)=\frac{E}{\exp{(\frac{E}{T})}-1 }$

To arrive at our final conclusion, we must show that $\zeta(E,T)/T$  is limited and converges uniformly to the function $\zeta_{\infty}=1$ in each closed interval $[a,b]$, $a,b\geq 0$ when $T\to \infty$. In other words, $\max_{x \in [a,b]}\left| 1-\zeta(E,T)/T\right|<\epsilon$  for each $\epsilon>0$ and $T$ sufficiently large, moreover, for all $E$, $\left|\zeta(E,T)/T\right|<C$, being $C$ an arbitrary positive constant.

  It is because if $\int_{a}^{b}\vartheta(E)dE$ is finite and absolutely integrable ($\int_a^b\left|\vartheta(E)\right|dE<\infty$), than $\left| \int_{a}^{b}\vartheta(E)-\frac{\zeta(E,T)}{T} \vartheta(E)dE\right|$ $\leq \int_{a}^{b}\left|\vartheta(E)-\frac{\zeta(E,T)}{T} \vartheta(E)\right|dE$ $\leq\epsilon\int_{a}^{b}\left|\vartheta (E)\right|dE$. We can let $b \to \infty$ in the inequality, because of the absolute convergence of the integral $\int_{a}^{\infty}\vartheta(E)dE$ by hypothesis and the upper bound on $\left|\zeta(E,T)/T\right|$:
\begin{equation*}
\begin{split}
&\quad \left| \int_{a}^{\infty}\vartheta(E)-\frac{\zeta(E,T)}{T} \vartheta(E)dE\right|\\
&\leq  \int_{a}^{\infty}\left|\vartheta(E)-\frac{\zeta(E,T)}{T} \vartheta(E)\right|dE\\
&\leq\epsilon\int_{a}^{b}\left|\vartheta (E)\right|dE+(1+C)\int_{b}^{\infty}\left|\vartheta (E)\right|dE \\
&\leq \epsilon\int_{a}^{b}\left|\vartheta (E)\right|dE+(1+C)\epsilon
\end{split}
\end{equation*}
To do this, we note that:
\begin{equation}\label{lim}
\lim_{E \to 0} \frac{E/T}{\exp(E/T)-1}=1
\end{equation}

We need to show that $\zeta(E,T)/T$ is a monotonically decreasing function of $E$:
\begin{gather*}
\frac{d}{dE}(\frac{E/T}{\exp(E/T)-1})=\frac{d}{TdE/T}(\frac{E/T}{\exp(E/T)-1})\\
=\frac{1}{T}\frac{d}{dy}\frac{y}{\exp(y)-1}\\
\frac{d}{dy}\frac{y}{\exp(y)-1}=\frac{1}{\exp(y)-1}-\frac{y\exp(y)}{(\exp(y)-1)^2}\\
=\frac{\exp(y)}{(\exp(y)-1)^2}\left(1-\frac{1}{\exp(y)}-y \right)<0
\end{gather*}
 for $y>0$.

To verify the last inequality, we proceed as follows:

For $y\geq1$, it is trivial.

Let us examine the case $0<y<1$.

We have that:
$$1-y-\frac{1}{exp(y)}= -\frac{y^2}{2!}+\frac{y^3}{3!}+\cdots$$

Because this power series is absolutely convergent ($\sum |a_nx^n|<\infty$), the summation order is irrelevant and we conclude that:
\begin{equation*}
\begin{split}
&1-y-\frac{1}{exp(y)}=\left(-\frac{y^2}{2!}+\frac{y^3}{3!}\right)+ \left(-\frac{y^4}{4!}+\frac{y^5}{5!}\right)+\\
&\cdots+ \left(-\frac{y^{2n}}{(2n)!}+\frac{y^{2n+1}}{(2n+1)!}\right)+\cdots
\end{split}
\end{equation*}
But,
\begin{equation*}
\begin{split}
&\left(-\frac{y^{2n}}{(2n)!}+\frac{y^{2n+1}}{(2n+1)!}\right)=y^{2n}\left( \frac{-1}{(2n)!} +\frac{y}{(2n+1)!} \right)\\
&<y^{2n}\left(\frac{-1}{(2n)!} +\frac{1}{(2n+1)!}  \right)<0
\end{split}
\end{equation*}
and we arrive at the desired inequality.

As a final step, because $\zeta(E,T)/T$ is monotonically decreasing, we have that:
$$ \max_{E\in [0,x]}\left|1-\frac{\zeta(E,T)}{T} \right|=1-\frac{\zeta(x,T)}{T}$$.
 $T\to \infty$, is equivalent to $\frac{x}{T}=y\to 0$, but, because of (\ref{lim}),   $$1-\frac{\zeta(x,T)}{T}\to 0$$
which concludes the proof.\Qed
\end{Proof}

\section{Proof of theorem 3}
\begin{theorem}
if $g\in G_1$, once we have $\lim_{T \to \infty }w>-1$, we will necessarily have $w(T)>-1$ for all $T$ if $g>-1/3$.
\end{theorem}
\begin{Proof}
We have that:
$$\frac{dw}{dT}=\frac{d\rho}{dT}\frac{1}{\rho}\left[\frac{dp/dT}{d\rho/dT}-\frac{p}{\rho} \right]$$

The first step is to show that $\frac{d\rho}{dT}>0$:
$$\rho=\int\frac{1}{\pi^2}\frac{E^3}{\exp(E/T)-1}\frac{1}{f^3}\left|g\right|dE$$

\begin{eqnarray*}
\frac{d\rho}{dT}&=&\int\frac{1}{\pi^2}\frac{\partial}{\partial T}\left( \frac{1}{\exp(E/T)-1}\right)\frac{E^3}{f^3}\left|g\right|dE\\
                &=&\int\frac{1}{\pi^2} \frac{\exp(E/T)}{(\exp(E/T)-1)^2}\frac{E}{T^2}\frac{E^3}{f^3}\left|g\right|dE>0
\end{eqnarray*}

The next step is to show that $p<0$ implies $\frac{dp}{dT}<0$:

It consists first in showing that $\frac{E\exp(E/T)}{\exp(E/T)-1}$ is a monotonically increasing function of $E$:
\begin{gather*}
\quad\frac{d}{dE}\left(\frac{E\exp(E/T)}{\exp(E/T)-1}\right)=\frac{d}{dy}\left(\frac{ye^y}{e^y-1}\right)\\
=\frac{e^y}{(e^y-1)^2}\left(-1-y+e^y\right)\\
=\frac{e^y}{(e^y-1)^2}\left(\frac{y^2}{2!}+\frac{y^3}{3!}+\cdots\right)>0\\
\end{gather*}
, for $y>0$.

We have that:

\begin{gather*}
3\frac{dp}{dT}=\int_{0}^{E_0} \frac{1}{\pi^2}\frac{e^{\frac{E}{T}}E}{(e^{E/T}-1)}\frac{1}{T^2}\frac{E^3}{(e^{E/T-1})}\frac{1}{f^3}dE\\
-\int_{E_0}^{\infty}\frac{1}{\pi^2}\frac{e^{\frac{E}{T}}E}{(e^{E/T}-1)}\frac{1}{T^2}\frac{E^3}{(e^{E/T-1})}\frac{1}{f^3}dE
\end{gather*}

if $p<0$, it implies:
$$\frac{dp}{dT}<\frac{E_0e^{\frac{E_0}{T}}}{e^{\frac{E_0}{T}-1}}\frac{p}{T^2}<0$$

We arrive at our desired result by showing that $\frac{dw}{dT}<0$ when $w=-1$:

Suppose that for some $T$ we have $w=-1$. Then for this $T$:
\begin{equation} \label{dwdT}
\frac{dw}{dT}=\frac{d\rho}{dT}\frac{1}{\rho}\left(\frac{|p|}{\rho}-\frac{\left|\frac{dp}{dT}\right|}{\frac{d\rho}{dT}}\right)
\end{equation}
\begin{gather*}
\frac{dp}{dT}=\frac13\int_{0}^{E_0} \frac{1}{\pi^2}\theta(E,T)\frac{E^3}{(e^{E/T-1})}\frac{1}{f^3}dE\\
-\frac13\int_{E_0}^{\infty}\frac{1}{\pi^2}\theta(E,T)\frac{E^3}{(e^{E/T-1})}\frac{1}{f^3}dE
\end{gather*}

$$\frac{d\rho}{dT}=\int_{0}^{\infty} \frac{1}{\pi^2}\theta(E,T)\frac{1}{T^2}\frac{E^3}{(e^{E/T-1})}\frac{1}{f^3}|g(E)|dE$$
$$\theta(E,T)=\frac{e^{\frac{E}{T}}E}{(e^{E/T}-1)}\frac{1}{T^2}$$

$$\frac{\left|\frac{dp}{dT}\right|}{\frac{d\rho}{dT}}=\frac{\frac{dp}{dT}/p}{\frac{d\rho}{dT}/\rho}.\frac{|p|}{\rho}$$

To calculate $\frac{d\rho}{dT}/\rho$, we make the change of variable:
$$x=\frac{\int_{0}^{E}\rho(\xi,T)d\xi}{\rho}$$

Because $\rho(E,T)\geq0$, the transformation $E\to x$ is invertible.
We conclude that:
$${\frac{d\rho}{dT}}\frac{1}{\rho}=\int_{0}^{1}\theta(E,T)dx$$

To do the same for ${\frac{dp}{dT}}/{p}$, an analogous transformation has some additional subtlety: $\rho(E,T)/g$ is not positive definite. Thus, the transformation:
$$y=\frac{\frac13\int_{0}^{E}\frac{\rho(\xi,T)}{g}d\xi}{p}$$
is not invertible.

Nevertheless, if $p<0$, there exists an $E_1^*>0$ such that $p=-\frac13\int_{E_1^*}^{\infty}\beta(\xi,T)d\xi$, where $\beta(E,T)=\frac{1}{\pi^2}\frac{E^3}{e^{E/T}-1}\frac{1}{f^3}$,
 and there exists an $E_2^*>0$ such that $\frac{dp}{dT}=-\frac13\int_{E_2^*}^{\infty}\theta(\xi,T)\beta(\xi,T)d\xi$ where $E_2^*<E_1^*$.

 We verify this:
\begin{gather*}
\int_{0}^{E_0}\theta(\xi,T)\beta(\xi,T)d\xi-\int_{E_0}^{E_1^*}\theta(\xi,T)\beta(\xi,T)d\xi\\
<\theta(E_0,T)\left[\int_{0}^{E_0}\beta(\xi,T)d\xi-\int_{E_0}^{E_1^*}\beta(\xi,T)d\xi\right]=0
\end{gather*}

 We used above the fact that $\theta(E,T)$ is an increasing function of $E$ and $p(T)= \frac13\left[\int_{0}^{E_0}\beta(\xi,T)d\xi-\int_{E_0}^{\infty}\beta(\xi,T)d\xi\right]<0$

 $$\frac{1}{p}\frac{dp}{dT}=\frac{-\frac{1}{3}\int_{E_2^*}^{E_1^*}\theta(\xi,T)\beta(\xi,T)d\xi-\frac13\int_{E_1^*}^{\infty}\theta(\xi,T)\beta(\xi,T)d\xi}{-\frac13\int_{E_1^*}^{\infty}\beta(\xi,T)d\xi}$$

We, instead, make the change $$y=\frac{\frac13\int_{E_1^*}^{E}\beta(\xi,T)d\xi}{|p|}$$
$$\Rightarrow\frac{1}{P}\frac{dP}{dT}=\tau+\int_{0}^{1}\theta(E,T)dy$$, where
$$\tau=\frac{-\frac{1}{3}\int_{E_2^*}^{E_1^*}\theta(\xi,T)\beta(\xi,T)d\xi}{-\frac13\int_{E_1^*}^{\infty}\beta(\xi,T)d\xi}$$

We then have:
$$\frac{\frac{1}{P}\frac{dp}{dT}}{\frac{1}{\rho}\frac{d\rho}{dT}}=\frac{\tau+\int_0^1\theta(E,T)dy}{\int_0^1\theta(E,T)dx}$$

Since $\tau\geq0$, we have from (\ref{dwdT}), that $dw/dT<0$ if
$$ \frac{\int_0^1\theta(E,T)dy}{\int_0^1\theta(E,T)dx}>1$$

For the sake of clarity of notation, let us denote by $\Psi$ the relation of $E$ and $y$, and by $\Phi$ the relation $E$ and $x$. We rewrite, the above expression as:
$$\frac{\int_0^1\theta(E,T)dy}{\int_0^1\theta(E,T)dx}=\frac{\int_0^1\theta(\Psi(z),T)dz}{\int_0^1\theta(\Phi(z),T)dz}$$

Again, we use the fact that $\theta$ is an increasing function of $E$ to assert that $\Psi(z)>\Phi(z)$ implies that $\theta(\Psi(z))>\theta(\Phi(z))$.

First, we note that $\Psi(0)=E^*_1$ and $\Phi(0)=0$. Since $\Psi-\Phi$ is continuous and $\Psi(0)-\Phi(0)>0$, $\Psi(z)-\Phi(z)<0$ for some $z>0$,
implies that there exists a $z^*$ such that $\Psi(z^*)-\Phi(z^*)=0$.

This is equivalent to
$$z^*=\frac13\int_{E^*_1}^{E^*}\beta(\xi,T)\frac{d\xi}{|p|}=\int_0^{E^*}\beta(\xi,T)|g(\xi)|\frac{d\xi}{\rho}$$

Which leads to:
$$1-z^*=\frac13\int_{E^*}^{\infty}\beta(\xi,T)\frac{d\xi}{|p|}=\int_{E^*}^{\infty}\beta(\xi,T)|g(\xi)|\frac{d\xi}{\rho}$$

However, by hypothesis, $|p|=\rho$ and $g(\xi)>-1/3$ so that
$$\frac13\int_{E^*}^{\infty}\beta(\xi,T)\frac{d\xi}{|p|}>\int_{E^*}^{\infty}\beta(\xi,T)|g(\xi)|\frac{d\xi}{\rho}$$

which is a contradiction that leads to the non-existence of $z^*$, which in turn results in $\Psi(z)>\Phi(z)$. This finishes the proof.\Qed
\end{Proof}
\section{proof of theorem 4}
\begin{theorem}
Suppose a functional sequence $g_N$, $N=1,2\cdots$ such that $($$g_N\in G_1$, $g_N>-\frac{1}{3}$ and $g_N(\xi_N)=0$$)$:
\begin{enumerate}[(I)]
\item $g_N(E)\geq 1-{E}/{\xi_N}$, $0\leq E\leq\xi_N$ for all $N$.\label{c1}\\
\item $|g_N(E)|<\epsilon_N$ for $E\geq\xi_N$.\label{c2}\\
\item $g_N<c_N$, $0<c_N<c$ and $(c_N-1)\xi_N<k$ for $E<\xi_N$.\label{c3}\\
\item $\epsilon_N\to 0$ when $N\to\infty$.\label{c4}\\
\item $\exists$ $\xi'$,  $0<\xi'<\xi_N$ such that $g_N<1+\alpha E$, $\alpha>0$ and $E\leq\xi'$\label{c5}\\
\end{enumerate}
then, $\lim_{N\to\infty}w_{N}^{\infty}=-\infty$.
\end{theorem}

\begin{Proof}

By using Eq. (\ref{WinftyS}) we claim that if we show that:

\begin{description}
  \item[ (a)] $\int_0^{\xi_N}\frac{E^2}{f_N^3}|g_N|dE/\int_0^{\xi_N}\frac{E^2}{f_N^3}dE<c$
  \item[ (b)] $\int_{\xi_N}^{\infty}\frac{E^2}{f_N^3}dE/\int_0^{\xi_N}\frac{E^2}{f_N^3}dE\to \infty$ when $N\to\infty$
  \item[ (c)] $\int_{\xi_N}^{\infty}\frac{E^2}{f_N^3}dE/\int_{\xi_N}^{\infty}\frac{E^2}{f_N^3}|g_N|dE\to \infty$ when $N\to\infty$
\end{description}

We conclude then the proof.

Indeed:

$$\lim_{N\to\infty}\frac{-\int_{\xi_N}^{\infty}\frac{E^2}{f_N^3}dE}{\int_0^{\xi_N}\frac{E^2}{f_N^3}dE-\int_{\xi_N}^{\infty}\frac{E^2}{f_N^3}dE}=\lim_{X\to\infty}\frac{-X}{1-X}=1$$
where $X=\int_{\xi_N}^{\infty}\frac{E^2}{f_N^3}dE/\int_0^{\xi_N}\frac{E^2}{f_N^3}dE$

It leads to:
\begin{eqnarray*}
\lim_{N\to\infty}w_N^{\infty}&=&\frac13\lim_{N\to\infty}\frac{-\int_{\xi_N}^{\infty}\frac{E^2}{f_N^3}dE}{\int_0^{\xi_N}\frac{E^2}{f_N^3}dE-\int_{\xi_N}^{\infty}\frac{E^2}{f_N^3}dE}\\
&\cdot&\frac{\int_0^{\xi_N}\frac{E^2}{f_N^3}dE-\int_{\xi_N}^{\infty}\frac{E^2}{f_N^3}dE}{\int_0^{\xi_N}\frac{E^2}{f_N^3}|g_N|dE+\int_{\xi_N}^{\infty}\frac{E^2}{f_N^3}|g_N|dE}\\
&=&\frac13\lim_{N\to\infty}\frac{-\int_{\xi_N}^{\infty}\frac{E^2}{f_N^3}dE}{\int_0^{\xi_N}\frac{E^2}{f_N^3}|g_N|dE+\int_{\xi_N}^{\infty}\frac{E^2}{f_N^3}|g_N|dE}\\
&=&\frac13\lim_{N\to\infty}\frac{-1}{\frac{\int_0^{\xi_N}\frac{E^2}{f_N^3}|g_N|dE}{\int_{\xi_N}^{\infty}\frac{E^2}{f_N^3}dE}+\frac{\int_{\xi_N}^{\infty}\frac{E^2}{f_N^3}|g_N|dE}{\int_{\xi_N}^{\infty}\frac{E^2}{f_N^3}dE}}\\
&<&\frac13\lim_{N\to\infty}\frac{-1}{\frac{c\int_0^{\xi_N}\frac{E^2}{f_N^3}dE}{\int_{\xi_N}^{\infty}\frac{E^2}{f_N^3}dE}+\frac{\int_{\xi_N}^{\infty}\frac{E^2}{f_N^3}|g_N|dE}{\int_{\xi_N}^{\infty}\frac{E^2}{f_N^3}dE}}\to-\infty
 \end{eqnarray*}
 It is because of (b) and (c). The $c$ factor appears due to the use of (a).

  let us show (a), (b) and (c).

(a)	Comes directly from (\ref{c3}).

(c)	Comes directly from (\ref{c2}) and(\ref{c4}).

Let us concentrate on showing (b):

From condition (\ref{c2}),

\begin{eqnarray*}
f_N(E)&=&f_N(\xi_N)e^{\int_{\xi_N}^{E}\frac{1-g(x)}{x}dx}\\
&<& f_N(\xi_N)e^{\int_{\xi_N}^{E}\frac{1+\epsilon_N}{x}dx}\\
&=&f_{N,\epsilon}(E)
\end{eqnarray*}
for $E>\xi_N$.

It implies:

$$\int_{\xi_N}^{\infty}E^2/f^3>\int_{\xi_N}^{\infty}E^2/f_{N,\epsilon}^3$$
where $f_{N,\epsilon}=f_N(\xi_N)\left(\frac{E}{\xi_N}\right)^{1+\epsilon_N}$.

Then,
$$\int_{\xi_N}^{\infty}E^2/f_{N,\epsilon}^3=\frac{1}{f_N(\xi_N)^3\xi_N^{-3}}\frac{1}{3\epsilon_N}$$

that in turn, implies that:

$$\frac{\int_{\xi_N}^{\infty}E^2/f_N^3}{\int_0^{\xi_N}E^2/f_N^3}>\frac{\int_{\xi_N}^{\infty}E^2/f_{N,\epsilon}^3}{\int_{0}^{\xi_N}E^2/f_N^3}$$

By using condition (\ref{c5}), the RHS results in being greater than:
$$\frac{\beta\int_{\xi_N}^{\infty}E^2/f_{N,\epsilon}^3}{\int_0^{\xi_N}E^2}=\frac{\beta}{f_N(\xi_n)^3\epsilon_N}$$

indeed:
for $E<\xi'$
\begin{eqnarray*}
g_N&<&1+\alpha E\\
\Rightarrow \exp(\int_0^{E}\frac{1-g_N}{E})&>&e^{-\alpha E}\\
\end{eqnarray*}
for $\xi'<E<\xi_N$ and using condition (\ref{c3}) again:

\begin{eqnarray*}
\Rightarrow f_N(E)&>&e^{-\alpha \xi'}\exp(\int_{\xi'}^{\xi_N}\frac{1-g_N}{E})\\
&>&e^{-\alpha \xi'}\exp(\int_{\xi'}^{\xi_N}\frac{1-c_N}{E})\\
&>&e^{-\alpha \xi'}\exp[\frac{1}{\xi'}\int_{\xi'}^{\xi_N}(1-c_N)]\\
&>&e^{-\alpha \xi'}e^{\frac{\kappa}{\xi'}}=\frac{1}{\beta}\\
\end{eqnarray*}

At last, to show that (b) is true requires showing that:

$f_N(\xi_N)\leq\tau$ for all $N$, because it implies that:
$$\frac{\beta}{f_N^3(\xi_N)\epsilon_N}\geq\frac{\beta}{\tau^3\epsilon_N}\to\infty$$

We have made use here of condition (\ref{c4}).

From condition (\ref{c1}):

$$f_N(\xi_N)=\exp(\int_0^{\xi_N}\frac{1-g_N}{E})\leq e$$ \Qed
\end{Proof}
\section{Proof of theorem 5}

\begin{theorem}
Suppose a functional sequence with $g_N(E)$ $($$g_N(E)\in G_1$ with $g_N(E)>-1/3$ and $g_N(\xi_N)=0$$)$ satisfying:
\begin{enumerate}[(I)]	
\item $g_N(E)<1+\alpha E$ for $E<\xi'$  and for all $N$ and $0<\alpha<\infty$\label{C1}\\
\item $g_N(E)>1-\beta E$ for some  $0<\beta<\infty$, for $E<\xi''<\xi'$ and for all $N$.\label{C2}\\
\item $g_N(E)<1/3-\epsilon$, for some $\epsilon>0$, $E\geq\xi'$ and for all $N$.\label{C3}\\
\item $\xi_N\to\infty$ when $N\to \infty$ \label{C4}\\
\item $g_N(E)<-1/3+\epsilon_N$ for $E>\xi_N+\Delta_N$, with $\epsilon_N\to 0$ and $\Delta_N\to0$ when $N\to \infty$\label{C5}\\
\end{enumerate}
, then, for sufficiently high $N$ we have $w_N^{\infty}>0$
\end{theorem}

\begin{Proof}

Condition (\ref{C5}) implies that $f(E)>f(\xi_N)\exp(\int_{\xi_N}^E\frac{1}{\varepsilon}d\varepsilon)$ for $\xi_N<E<\xi_N+\Delta_N$ and $f(E)>f(\xi_N+\Delta_N)\exp(\int_{\xi_N+\Delta_N}^E\frac{4/3-\epsilon_N}{\varepsilon}d\varepsilon)$ for $E>\xi_N+\Delta_N$. Then:
 \begin{equation}\label{OtherSide}
 \begin{split}
 \int_{\xi_N}^{\infty}\frac{E^2}{f^3}<&\int_{\xi_N}^{\xi_N+\Delta_N}\frac{E^2}{f(\xi_N)^3\left(\frac{E}{\xi_N}\right)^{3}}dE\\
 +&\int_{\xi_N+\Delta_N}^{\infty}\frac{E^2}{f(\xi_N+\Delta_N)^3\left(\frac{E}{\xi_N+\Delta_N}\right)^{4-3\epsilon_N}}dE\\
  =&\frac{{\xi_N}^3}{f(\xi_N)^3}\ln\left(\frac{\xi_N+\Delta_N}{\xi_N}\right)+\frac{(\xi_N+\Delta_N)^{3}}{f(\xi_N+\Delta_N)^3}\frac{1}{1-3\epsilon_N}\\
  \to &\frac{\xi_N^3}{f(\xi_N)^3}\mbox{, when $N\to\infty$}\\
 \end{split}
\end{equation}

Additionally, because for $\xi_N<E<\xi_N+\Delta_N$ we have $g_N(E)>-1/3$ than:
\begin{equation*}
\begin{split}
\frac{(\xi_N+\Delta_N)^3}{f(\xi_N+\Delta_N)^3}&=\exp(3\ln(\xi_N+\Delta_N)\\
                 &-3\ln f(\xi_N)-3\int_{\xi'}^{\xi_N+\Delta_N}\frac{1-g_N(\varepsilon)}{\varepsilon}d\varepsilon)\\
                 &=\exp(3\ln \xi_N-3\ln f(\xi_N)\\
                 &-\int_{\xi'}^{\xi_N+\Delta_N}\frac{-3g_N(\varepsilon)}{\varepsilon}d\varepsilon)\\
                 &>\frac{(\xi_N)^3}{f(\xi_N)^3}\left(\frac{\xi_N}{\xi_N+\Delta_N}\right)
\end{split}
\end{equation*}

That in turn, implies that
\begin{gather}\label{greater_than_one}
\frac{{\xi_N}^3}{f(\xi_N)^3}\ln\left(\frac{\xi_N+\Delta_N}{\xi_N}\right)+\frac{(\xi_N+\Delta_N)^{3}}{f(\xi_N+\Delta_N)^3}\frac{1}{1-3\epsilon_N}\nonumber\\
>\frac{{\xi_N}^3}{f(\xi_N)^3}\ln\left(\frac{\xi_N+\Delta_N}{\xi_N}\right)+\frac{(\xi_N)^{3}}{f(\xi_N)^3}\left(\frac{\xi_N}{\xi_N+\Delta_N}\right)\frac{1}{1-3\epsilon_N}\nonumber\\
=\frac{{\xi_N}^3}{f(\xi_N)^3}\ln X+\frac{\xi_N^{3}}{f(\xi_N)^3}\frac{1}{X}\frac{1}{1-3\epsilon_N}\nonumber\\
>\frac{{\xi_N}^3}{f(\xi_N)^3}\ln X+\frac{\xi_N^{3}}{f(\xi_N)^3}\frac{1}{X}\nonumber\\
>\frac{{\xi_N}^3}{f(\xi_N)^3}\nonumber\\
\end{gather}

Above, we have used the fact that $\ln X+ 1/X$ is $1$ for $X=1$ and $\frac{d}{dX}\left(\ln X+ 1/X\right)=\frac{1}{X}-\frac{1}{X^2}>0$ for $X>1$.

Condition (\ref{C3}) implies that $\frac{d}{dE}\frac{E^2}{f^3}<0$ for $E>\xi'$:
 \begin{equation*}
 \begin{split}
 \frac{d}{dE}\frac{E^2}{f^3}=&\frac{E^2}{f^3(\xi')\exp(3\int_{\xi'}^{E}\frac{1-g(\varepsilon)}{\varepsilon}d\varepsilon)}\\
 =&\frac{d}{dE}\exp(2\ln E-3\ln( f(\xi'))-3\int_{\xi'}^{E}\frac{1-g_N(\varepsilon)}{\varepsilon}d\varepsilon)\\
 =&\frac{d}{dE}\exp(2\ln \xi' -3\ln( f(\xi'))-\int_{\xi'}^{E}\frac{1-3g_N(\varepsilon)}{\varepsilon}d\varepsilon)\\
 =&-\frac{d}{dE}\int_{\xi'}^{E}\frac{1-3g_N(\varepsilon)}{\varepsilon}d\varepsilon \frac{E^2}{f^3}
 \end{split}
 \end{equation*}
then: $$\int_{\xi'}^{\xi_N}\frac{E^2}{f^3}dE>\int_{\xi'}^{\xi_N}\frac{\xi_N^2}{f(\xi_N)^3}dE$$

For the other side:
$$\frac{\xi_N^3}{f^3(\xi_N)}=\int_{\xi'}^{\xi_N}\frac{\xi_N^2}{f(\xi_N)^3}+\frac{\xi_N^2}{f(\xi_N)^3}\cdot\xi'$$

Provided that $0<\Gamma_1<\frac{\xi'^2}{f(\xi')^3}<\Gamma_2$, condition (\ref{C3}) implies that $\frac{E^2}{f(E)^3}<\Gamma_2\frac{\xi'^{3\epsilon}}{E^{3\epsilon}}$ for $E>\xi'$, that means that $\frac{E^2}{f(E)^3}$ can be made arbitrarily small for sufficiently large $E$. This, in turn, implies that $\frac{\xi_N^2}{f(\xi_N)^3}\xi'$ can be made arbitrarily small by condition (\ref{C4}). For the other side, $\frac{\xi_N^3}{f(\xi_N)^3}>\Gamma_1\frac{\xi'}{\xi_N}$ that cannot be made arbitrarily small.

By condition (\ref{C1}) we can make $\Gamma_2={\xi'^2}/{e^{-\alpha\xi'}}$, and by condition (\ref{C2}) we can make $\Gamma_1={\xi'^2}/{e^{\beta\xi''}\frac{\xi''}{\xi'}}$.

Because $\int_{\xi'}^{\xi_N}\frac{E^2}{f^3}dE/\frac{\xi_N^2}{f(\xi_N)^3}({\xi_N}-{\xi'})>1$ and
$\frac{\xi_N^2}{f(\xi_N)^3}{\xi_N}>\Gamma_1\xi'>0$ for all N, while $\frac{\xi_N^2}{f(\xi_N)^3}\xi'$ is arbitrarily small. We have that:

$$\int_{0}^{\xi_N}\frac{E^2}{f^3}dE>\frac{\xi_N^3}{f(\xi_N)^3}\mbox{, for sufficiently high $N$}$$

At this moment, we can not conclude that $\int_{0}^{\xi_N}\frac{E^2}{f^3}>\int_{\xi_N}^{\infty}\frac{E^2}{f^3}$ for sufficiently large $N$. It is because the estimate (\ref{OtherSide}) says only that $\int_{\xi_N}^{\infty}\frac{E^2}{f^3}$ is less than something always greater than $\frac{\xi_N^3}{f(\xi_N)^3}$ (By \ref{greater_than_one}), then it can be greater than this.

To arrive at our final conclusion, we must show that $\int_{\xi'}^{\xi_N}\frac{E^2}{f^3}dE/ \frac{\xi_N^2}{f(\xi_N)^3}({\xi_N}-{\xi'})>1+\delta$ with $\delta>0$ for sufficiently large $N$, since by estimate (\ref{OtherSide}), we have that $\left(\frac{{\xi_N}^3}{f(\xi_N)^3}\ln\left(\frac{\xi_N+\Delta_N}{\xi_N}\right)+\frac{(\xi_N+\Delta_N)^{3}}{f(\xi_N+\Delta_N)^3}\frac{1}{1-3\epsilon_N}\right)/\frac{\xi_N^3}{f(\xi_N)^3}$ is always greater and arbitrarily close to 1.

This being true, we have that:
\begin{gather*}
\frac{\int_{0}^{\xi_N}\frac{E^2}{f^3}dE-\int_{\xi_N}^{\infty}\frac{E^2}{f^3}dE}{{\xi_N^3}/{f(\xi_N)^3}}\\
>\frac{\int_{0}^{\xi_N}\frac{E^2}{f^3}dE
-\left(\frac{{\xi_N}^3}{f(\xi_N)^3}\ln\left(\frac{\xi_N+\Delta_N}{\xi_N}\right)+\frac{(\xi_N+\Delta_N)^{3}}{f(\xi_N+\Delta_N)^3}\frac{1}{1-3\epsilon_N}\right)}{{\xi_N^3}/{f(\xi_N)^3}}>\delta
\end{gather*}
, for sufficiently large N, which concludes the proof by (\ref{WinftyS}).

By condition (3) we have that when $N\to\infty$:
\begin{equation}\label{epsilon3}
\frac{\int_{\xi'}^{\xi_N}\frac{E^2}{f^3}dE}{\frac{\xi_N^2}{f(\xi_N)^3}(\xi_N-\xi')}>\frac{1}{1-3\epsilon}
\end{equation}

It is because if $\varrho(E)/\varrho(\xi_N)>\varsigma(E)/\varsigma(\xi_N)$ for all $E\in(\xi',\xi_N)$ then:
\begin{equation}\label{varrho}
\frac{\int_{\xi'}^{\xi_N}\varrho(E)dE}{\varrho(\xi_N)(\xi_N-\xi')}> \frac{\int_{\xi'}^{\xi_N}\varsigma(E)dE}{\varsigma(\xi_N)(\xi_N-\xi')}
\end{equation}
Condition (3) puts a inferior limit on this ratio: setting $\varrho(E)=\frac{E^2}{f^3}$, we have that that $\frac{\varrho(E)}{\varrho(\xi_N)}>\left(\frac{\xi_N}{E}\right)^{3\epsilon}$.
It concludes the proof when we calculate the right side of (\ref{varrho}) for  $\frac{\xi'^2}{f(\xi')^3}\frac{\xi'^{3\epsilon}}{E^{3\epsilon}}$ take the $N\to\infty$ and obtain (\ref{epsilon3}).
\Qed
\end{Proof}
\section{Proof of theorem 6}
\begin{lemma}

Let us consider an expression like
\begin{equation}\label{continuousIntegral}
\int_0^{\xi_N}\varrho(E,\alpha)dE-\int_{\xi_N}^{\infty}\varrho(E,\alpha)dE\mbox{,}
\end{equation}
with $\xi_N$ a constant satisfying $\xi_N\geq0$ or a continuous function of $\alpha$, $\alpha\in[\alpha_1,\alpha_2]$ with $\xi_N>0$ for all $\alpha$ and suppose that $\varrho(E,\alpha)$ is continuous$($ in the following sense:
for each $\epsilon>0$ there exists a $\delta>0$ such that $|\varrho(E_1,\alpha_1)-\varrho(E_0,\alpha_0)|<\epsilon$ if $||(E_1,\alpha_1)-(E_0,\alpha_0)||<\delta$ $($$||x||$ denotes the usual Euclidian norm$))$. Let us suppose additionally that for each $\epsilon>0$ there exists an $\xi_{\epsilon}$ such that
\begin{equation}\label{xie}
\int_{\xi_{\epsilon}}^{\infty}\varrho(E,\alpha)dE<\epsilon\mbox{ for all $\alpha$.}
\end{equation}
 Then the expression (\ref{continuousIntegral}) is continuous with respect to the $\alpha$ parameter.
\end{lemma}

\begin{Proof}
Let us begin with the first integral in the case in which $\xi_N$ is constant:
To each $(E_0,\alpha_0)$ with fixed $\alpha_0$ and $E_0\in[0,\xi_n]$, inside the open ball defined by $||(E,\alpha)-(E_0,\alpha_0)||<\delta$ such that $|g(E,\alpha)-g(E_0,\alpha_0)|<\epsilon$ there is an open box $|E-E_0|<\delta _{B}$ and $|\alpha-\alpha_0|<\delta_{B}$. The union of such boxes is an open cover of the closed interval $[0,\xi_N]$.

It is a basic result in topology that on the real line, every bounded and closed set is compact, and, by definition, to every open cover of a compact set there exists a finite sub-cover. Then, a finite number of them cover the compact interval. Of this finite set, choose a the greatest $\delta_{B}$ and denote it $\delta_{M}$. It implies that $|g(E,\alpha)-g(E,\alpha_0)|<\epsilon$ if $|\alpha-\alpha_0|<\delta_{M}$. In turn, implies that $|\int_{0}^{\xi_N}\varrho(E,\alpha)dE-\int_{0}^{\xi_N}\varrho(E,\alpha_0)|<\epsilon\xi_N$ if $|\alpha-\alpha_0|<\delta_{M}$, proving the continuity of the first integral.

To prove the continuity of the second, consider $\delta_{B}$ such that $|\alpha-\alpha_0|<\delta_{B}$ implies $|\int_{\xi_N}^{\xi_{\epsilon}}\varrho(E,\alpha)dE-   \int_{\xi_N}^{\xi_{\epsilon}}\varrho(E,\alpha_0)dE|<\epsilon$, then

\begin{gather*}
|\int_{\xi_N}^{\infty}\varrho(E,\alpha)dE-\int_{\xi_N}^{\infty}\varrho(E,\alpha)dE|\\
=|\int_{\xi_N}^{\xi_{\epsilon}}\varrho(E,\alpha)dE-\int_{\xi_N}^{\xi_{\epsilon}}\varrho(E,\alpha_0)dE\\
+\int_{\xi_{\epsilon}}^{\infty}\varrho(E,\alpha)dE-\int_{\xi_{\epsilon}}^{\infty}\varrho(E,\alpha_0)dE|\\
\leq|\int_{\xi_N}^{\xi_{\epsilon}}\varrho(E,\alpha)dE-\int_{\xi_N}^{\xi_{\epsilon}}\varrho(E,\alpha_0)dE|\\
+|\int_{\xi_{\epsilon}}^{\infty}\varrho(E,\alpha)dE|+|\int_{\xi_{\epsilon}}^{\infty}\varrho(E,\alpha_0)dE|\leq 3\epsilon
\end{gather*}
when $|\alpha-\alpha_0|<\delta_{B}$.
To extend the last result to the case where $\xi_N$ is variable, consider only that $\xi_N$ is a continuous function of $\alpha$ and $\xi_N(\alpha)>0$ for all $\alpha$. Then make a variable change defined by $\varepsilon=E\frac{\xi_N(\alpha_1)}{\xi_N(\alpha)}$. We work instead with
$$\int_{0}^{\xi_N(\alpha_1)} \upsilon(\varepsilon,\alpha)\frac{\xi_N(\alpha)}{\xi_N(\alpha_1)}d\varepsilon-\int_{\xi_N(\alpha_1)}^{\infty} \upsilon(\varepsilon,\alpha)\frac{\xi_N(\alpha)}{\xi_N(\alpha_1)}d\varepsilon$$

The validity of condition (\ref{xie}) for $\varrho(E,\alpha)$ implies the validity for $\upsilon(\varepsilon,\alpha)\frac{\xi_N(\alpha)}{\xi_N(\alpha_1)}$ since $[\alpha_1,\alpha_2]$ is compact.

Since $\xi_N(\alpha)$ is continuous, it reaches its maximum at $\alpha_{max}$ and minimum at $\alpha_{min}$ on the compact interval $[\alpha_1,\alpha_2]$. Then:
\begin{gather*}
\int_{\varepsilon>\frac{\xi_N(\alpha_1)}{\xi_N(\alpha_{min})}\xi_{\epsilon}}\upsilon(\varepsilon,\alpha)\frac{\xi_N(\alpha)}{\xi_N(\alpha_1)}d\varepsilon\\
<\int_{\varepsilon>\frac{\xi_N(\alpha_1)}{\xi_N(\alpha)}\xi_{\epsilon}}\upsilon(\varepsilon,\alpha)\frac{\xi_N(\alpha)}{\xi_N(\alpha_1)}d\varepsilon<
\epsilon
\end{gather*}\Qed

\end{Proof}

\begin{theorem}
Suppose $g(E,\alpha)$ a limited $(|g(E,\alpha)|<C$ for all $\alpha$ and all $E)$ and differentiable function such that $g(E,\alpha)\in G_1$ for all $\alpha\in[\alpha_1,\alpha_2]$ $($and by consequence of definition of $G_1$, $\frac{dg(E,\alpha)}{dE}<0$ when $E=\xi_N(\alpha)$ such that $g(E,\xi_N(\alpha))=0$ $)$, then:
\begin{equation}\label{goal}
\frac{1}{3}\frac{\int_0^{\xi_N(\alpha)}E^2/f(E,\alpha)^3dE-\int_{\xi_N(\alpha)}^{\infty}E^2/f(E,\alpha)^3dE}{\int_0^{\infty}\left| g(E,\alpha) \right|E^2/f(E,\alpha)^3dE}
\end{equation}
is continuous with respect to $\alpha$ parameter.
\end{theorem}

\begin{Proof}
$g(E,\alpha)$ to be continuous as a function of $E$ and $\alpha$ in the sense discussed, assures the continuity of $f(E,\alpha)$ (since $f(E,\alpha)=e^{\int_0^E\frac{1-g(\varepsilon,\alpha)}{\varepsilon}d\epsilon}$).

To assure the continuity of $\xi_N(\alpha)$ such that $g(\xi_N(\alpha),\alpha)=0$ as a function of $\alpha$, it suffices that $\frac{dg(E,\alpha)}{dE}<0$ when $E=\xi_N(\alpha)$. It is an immediate application of the inverse function theorem:

Define The map $\Phi:[0,\infty)\times[\alpha_1,\alpha_2]$ by $\Phi(E,\alpha)=(g(E,\alpha),\alpha)$. Then, the differential of $\Phi$ is invertible on $(0,\alpha)$, that implies that $\Phi$ is a diffeomorphism on a neighborhood of this point, that in turn, implies that locally  $\xi_N(\alpha)$ is a differentiable function of $\alpha$ for all $\alpha$.

Suppose that for all $\alpha$ we have $g(E,\alpha)\in G$, then, there exists $g_{u}(E)\in G$ such that $g(E,\alpha)<g_{u}(E)$ for all $\alpha$. If this is true, then :
$$\int_{E>\xi_{\epsilon}}\frac{E^2}{f_{u}(E)^3}>\int_{E>\xi_{\epsilon}}\frac{E^2}{f(E,\alpha)^3}\mbox{ for all $\alpha$,}$$
satisfying condition (\ref{xie}), since the first integral was proved to exist in Appendix B, proving the continuity of the numerator. It finishes the proof, since $|g(E,\alpha)|<C$ that implies:
$$\int_{E>\xi_{\epsilon}}C\frac{E^2}{f_{u}(E)^3}>\int_{E>\xi_{\epsilon}}g(E,\alpha)\frac{E^2}{f(E,\alpha)^3}\mbox{ for all $\alpha$,}$$
, which assures the continuity of the denominator of (\ref{goal}).

We show that there exists $g_{u}(E)$ by showing that there exists $E_1$ such that $\max_{\alpha}g(E,\alpha)<-\epsilon$ for $E>E_1$. Then, define $g_{u}(E)$ with $g_{u}(\xi)=0$ for some $\xi>E_1$, $g_{u}(E)>-\epsilon$ for $E>\xi$ and $g_{u}(E)\geq\max_{\alpha}g(E,\alpha)$ for $E<\xi$.

To show that this $E_1$ in fact does exist proceed as follow: Suppose that for all $\alpha$ we have $g(E,\alpha)\in G$. Since $g(E,\alpha)$ is continuous on $\alpha$ that lies on the compact interval $[\alpha_1,\alpha_2]$ for all $E$, there exists $\max_{\alpha\in[\alpha_1,\alpha_2]}g(E,\alpha)=g(E,\alpha^E_{max})$. Suppose that $\lim_{E\to\infty}g(E,\alpha^E_{max})=0$, then, there exists an $\alpha_{max}$ sequence denoted by $\alpha_{max}^{E_1},\alpha_{max}^{E_2}\mbox{,...}$ such that $g(E_N,\alpha_{max}^{E_N})\to 0$.

But, on a compact interval, all infinite sequences have convergent subsequences. Let us suppose that the $\alpha_{max}^{N}$ sequence is that sequence and has a limit $\alpha_{max}^{\infty}$. But $\lim_{N\to\infty}|g(E_N,\alpha_{max}^{N})-g(E_N,\alpha_{max}^{\infty})|=0$, then $\lim_{N\to\infty}g(E_N,\alpha_{max}^{\infty})=0$. It implies that for $\alpha_{max}^{\infty}$ there is no $\epsilon>0$ such that $g(E,\alpha_{max}^{\infty})>-\epsilon$ for some $E>E_1$. Then, it does not belong to $G$, a contradiction. \Qed
\end{Proof}


\end{document}